\documentclass[aps,prx,superscriptaddress,twocolumn,floatfix,longbibliography]{revtex4-2} 

\usepackage[final]{graphicx}    
\graphicspath{{figures_paper/}}
\usepackage{soul}
\usepackage[utf8]{inputenc}
\usepackage{natbib}
\usepackage{xcolor}
\usepackage{amsfonts}
\usepackage{amsmath}
\usepackage{bm}
\usepackage{hyperref}
\hypersetup{colorlinks=true}

\definecolor{asparagus}{rgb}{0.53, 0.66, 0.42}

\newcommand{\Ket}[1]{\left|#1\right>}
\newcommand{\Bra}[1]{\left<#1\right|}

\newcommand{\mbf}[1]{\mathbf{#1}}

\begin{document}

\title{Floquet time crystals in driven spin systems with all-to-all $p$-body interactions}
\author{Manuel H. Muñoz-Arias}
\email{mhmunoz@unm.edu}
\author{Karthik Chinni}
\author{Pablo M. Poggi}
\email{ppoggi@unm.edu}
\affiliation{Center for Quantum Information and Control, Department of Physics and Astronomy, University of New Mexico, Albuquerque, New Mexico 87131, USA}
\begin{abstract}
We show the emergence of Floquet time crystal (FTC) phases in the Floquet dynamics of periodically driven $p$-spin models, which describe a collection of spin-1/2 particles with all-to-all $p$-body interactions. Given the mean-field nature of these models, we treat the problem exactly in the thermodynamic limit and show that, for a given $p$, these systems can host various robust time-crystalline responses with period $nT$, where $T$ is the period of the drive and $n$ an integer between 2 and $p$. In particular, the case of four-body interactions ($p=4$) gives rise to both a usual period-doubling crystal, and also a novel period-quadrupling phase. We develop a comprehensive framework to predict robust subharmonic response in classical area-preserving maps, and use this as a basis to predict the occurrence and characterize the stability of the resulting mean-field FTC phases in the quantum regime. Our analysis reveals that the robustness of the time-crystal behavior is reduced as their period increases, and establishes a connection between the emergence of time crystals, described by eigenstate ordering and robust subharmonic response, and the phenomenology of excited state and dynamical quantum phase transitions. Finally, for the models hosting two or more coexisting time crystal phases, we define protocols where the periodic subharmonic response of the system can be varied in time via the non-periodic modulation of an external control parameter.
\end{abstract}

\date{\today}
\maketitle


\section{Introduction}
\label{sec:intro} 
While historically the study of many-body quantum systems has been focused on its equilibrium and near-equilibrium properties, the emergence of prototypical quantum simulation platforms in recent years, such as ultracold atoms in optical lattices, trapped ions, and superconducting circuits, have opened the door to the study of various few- and many-body out-of-equilibrium phenomena~\cite{Polkovnikok2011}. These include quantum quenches~\cite{Ebadi2021} and closed-system thermalization~\cite{Kaufman2016}, dynamical phase transitions~\cite{Zhang2017}, and quantum many-body scars~\cite{Bluvstein2021}, among others. The interest has also been extended to driven systems which by definition lack an equilibrium regime but can, in some situations, lead to the emergence of nonequilibrium phases of matter, described by the robust nonstationary behavior of order parameters for generic initial states~\cite{Else2020}. One of the most prominent examples in this area is that of Floquet time crystals (FTCs)~\cite{else2016,Khemani2016}, which are systems described by a time-periodic Hamiltonian $H(t)$ and thus possess a discrete time-translation symmetry $H(t+T)=H(t)$, and in which the dynamics of generic observables display robust periodic oscillations with a period which is an integer multiple of $T$ (but not $T$). In this way, the behavior of the system is seen to break the aforementioned symmetry of the Hamiltonian. FTCs describe a scenario in which the interplay between the many-body interactions and the drive stabilizes the response of the system to show long-lived oscillations, instead of leading to its relaxation. 

Even though a general FTC can have in principle a period of $nT$, the case of $n=2$ has remained the most common case in previous works, with a plethora of theoretical as well as experimental studies related to this particular case~\cite{russomanno2017,Zhang2017,Ippoliti2021,Mi2021,Randall2021}. This type of FTC is particularly common in systems of driven interacting spin-$1/2$ particles which naturally present a $\mathbb{Z}_2$ symmetry. In this context, we seek to develop a framework for obtaining FTCs with a higher-order periodic response (i.e., a ``large-period" FTC), understand their robustness and physical origin, as well as their potential experimental implementation. In previous works, different alternatives have been proposed to go beyond period $2T$-FTCs~\cite{choi2017,Yao2017,Pizzi2019,Choudhury2021}. Many of these are based on using different types of interacting subsystems, such as bosons or higher-spin systems, which do not present the natural $\mathbb{Z}_2$ symmetry that leads to $2T$-FTCs. More recently, other proposals have been put forward to systematically construct $nT$-FTCs, including systems inspired in quantum error correcting codes \cite{bomantara2021} and clock models~\cite{Surace2019} (see also Ref. \cite{pizzi2021} described below).

In this work we show that periodically-driven spin models with all-to-all homogeneous $p$-body interactions can host robust time crystalline behavior with $nT$ periodic response where $n$ is as small as 2, but can be as high as $p$ in general. In these models the mean-field limit describes the dynamics of the system exactly in the thermodynamic limit, and so they constitute an example of a \textit{mean-field} FTC~\cite{Else2020,Natsheh2021,Natsheh2021b,Collado2021}. We show that mean-field FTC phases can be thoroughly characterized by the area-preserving map (APM) describing the dynamics of the collective spin in the mean-field limit. Using tools from dynamical systems theory, we provide a formal description of defining features of FTCs, such as subharmonic response and eigenstate order, in terms of classical precursors such as phase-space resonances and bifurcations. We then use this description to predict the occurrence of FTCs in the quantum regime, and to construct appropriate metrics to characterize them. 

Our work significantly expands previous studies on mean-field FTC phases in spin systems with infinite range two-body interactions~\cite{russomanno2017,pizzi2021,nurwantoro2019}, a scenario which naturally leads to lack of ergodicity even in clean systems without disorder or many-body localization. In particular, Ref.~\cite{pizzi2021} described the emergence of FTCs in these systems by identifying the periodic orbits of the classical area-preserving map describing the dynamics in the thermodynamic limit, and found subharmonic response with different frequencies depending on the system parameters. In this work we show that by considering the generalized case of $p$-body interactions, we can systematically construct robust higher-order FTCs. This reveals a connection between FTCs and the physical complexity of the system as described by degree of the interactions, or equivalently, by a higher degree of nonlinearity.

Furthermore, in this work we present new methods for diagnosing and controlling FTCs that reveal connections with well-known physical mechanisms, and which could find applications beyond the case of mean-field FTCs. Particularly, we show that eigenstate ordering in the Floquet operator can be diagnosed via the spectral statistics of an appropriate power of this operator, which in turn allows us to build a connection between eigenstate ordering and excited state quantum phase transitions~\cite{Cejnar2021}. We also show that the transition from a non-FTC to an FTC phase (or vice-versa) can be treated as a dynamical quantum phase transition~\cite{Zunkovic2018,Heyl2018rev}, and although the typical dynamical order parameters fail at detecting it, certain higher order correlation functions can be used to characterize the dynamical phase. Finally, we show that, as a consequence of the multi-body interactions, these models host FTCs of different orders for the same class of initial states, and that the order of the FTC can be tuned by a single parameter which can be physically regarded as an external field. As a result, we propose and numerically study a time crystal switching protocol whereby quenching a parameter in the Hamiltonian allows us to change between FTC phases with different periodic response.
 
The remainder of the manuscript is organized as follows. In Sec. \ref{sec:clean_and_drive} we review the definition of FTCs and introduce the system under study, the family of $p$-spin Hamiltonians. We also describe the driving protocol that maps the time-independent Hamiltonian into a Floquet system. In Sec. \ref{sec:reso_reso_cond_bifus} we analyze the mean-field limit of these driven models, and describe the mechanisms leading to robust subharmonic response in terms of resonances and signatures of structural changes in phase space (bifurcations).

In Sec. \ref{sec:characterization} we introduce the quantities we use to characterize the emergent FTC behavior in this system, both for finite sizes and in the thermodynamic limit, and discuss connections with notions of out-of-equilibrium quantum phase transitions. In Sec. \ref{sec:emergence_ftc} we present extensive numerical calculations providing evidence for the existence and robustness of the FTC phases in driven $p$-spin systems.
In Sec. \ref{sec:time_crystal_switching} we introduce the idea of \textit{time crystal switching}, a control protocol aimed at models with two or more coexisting FTC phases, \textit{i.e} $p \ge 4$, which allow us to switch the periodic response of the system and thus dynamically modulate between different FTC phases. Finally in Sec. \ref{sec:final_remarks} we present our conclusions and discuss potential future directions related to this work.

\section{Floquet time crystals, model and driving protocol}
\label{sec:clean_and_drive}
\subsection{Summary of Floquet time crystals}
\label{subsec:brief_on_FTC}
A periodically driven Hamiltonian system exhibits discrete time-translation symmetry, that is, invariance of the Hamiltonian at time intervals separated by one period of the drive, $T$, $\hat{H}(t + T) = \hat{H}(t)$. A FTC is an out-of-equilibrium phase of matter emerging as a consequence of a physical observable breaking the discrete time translation symmetry of a many-body Hamiltonian~\cite{else2016,Else2020}. More precisely, the FTC phase can be defined by considering a \textit{class} of initial states $\{|\psi\rangle\}$ and a generic choice of observable $\hat{O}$, such that the time-dependent expectation value in the limit of large system size $N$, given by  
\begin{equation}
    f_O(t)=\lim\limits_{N\rightarrow \infty} \Bra{\psi(t)}\hat{O}\Ket{\psi(t)},
\end{equation} 
 satisfies the following conditions:  \cite{russomanno2017,huang2018}
\begin{enumerate}
    \item \textbf{Time-translation symmetry breaking}: $f_O(t + T)\neq f_O(t)$ while $\hat{H}(t + T) = \hat{H}(t)$
    \item \textbf{Rigidity}: $f_O(t)$ has a fixed oscillation period, without needing to fine-tune parameters in $\hat{H}$.
    \item \textbf{Persistence}: the oscillations of $f_O(t)$ persist for an infinitely long time
\end{enumerate}
The conditions (1)-(3) imply that an FTC phase is not possible for generic chaotic systems, in which diffusion and equilibration typically preclude the existence of long-lived oscillations. Instead, FTC phases are expected to exist either in disordered systems (where diffusion can be suppressed by many-body localization~\cite{else2016,Khemani2016} or by other means~\cite{choi2017,ho2017}) or in certain clean, integrable systems with highly regular motion. Clean Floquet time crystals have been studied in~\cite{huang2018} for a Bose-Hubbard ladder, in~\cite{russomanno2017} for the Lipkin-Meshkov-Glick model, and in~\cite{Surace2019} for a family of clock models. See also~\cite{sacha2015} for a precursor of these studies.

The emergence of an FTC implies an structural change in the nature of the Floquet states, which are the eigenstates of the unitary evolution operator $\hat{U}_{F}$ corresponding to $\hat{H}$ after one driving period. A key signature of this structural change is the emergence of \textit{eigenstate ordering}~\cite{Else2020}, that is, the Floquet states become cat-like states, \textit{i.e.}, superpositions of macroscopically distinct states~\cite{else2016}. This is reflected in the reorganization of the Floquet spectrum, where an extensive portion of it is composed of groups of Floquet phases with spacing on the unit circle by an angle of $2\pi/q$, with $q$ and integer equal to the number of states entering in the superposition. As a consequence, the Floquet spectrum of $\hat{U}_F^q$ will be composed of clusters of $q$-fold degenerated Floquet phases, as all the Floquet phases in one of the groups seen in the spectrum of $\hat{U}_F$ will collapse to the same value. Eigenstate ordering has been recently proposed as the fundamental characteristic of an FTC phase~\cite{Else2020}, with the typical subharmonic response as its dynamical manifestation.

\subsection{$p$-spin models}  
\label{subsec:magnetic_model}
In the following we focus on a class of interacting magnetic systems called $p-$spin models~\cite{Bapst2012,Jorg2010,Matsuura2017}. Consider a system of $N$ spin-1/2 particles with $p$-body, $p\ll N$, Ising-like interactions, and in presence of an external homogeneous magnetic field, described by a time-independent Hamiltonian $H_p$. We will consider interactions to be homogeneous and of infinite range, \textit{i.e}, all-to-all, as to avoid heating to a featureless ``thermal'' phase when the external driving is included~\footnote{In presence of short-range interactions, the full Hilbert space of the $N$ spin-$1/2$ particles is considered, and typically one requires the addition of disorder to induce localization and salvage some region of parameter space from the deleterious effects of heating~\cite{else2016,Khemani2016,vonKeyserlingk2016}}. The interaction Hamiltonian is permutationally symmetric and given an initial state in the symmetric subspace of the $N$ spin-$1/2$ particles, any time evolution will be restricted to that space. The Hamiltonian  of the resulting family of $p$-spin models is
\begin{equation}
\label{eqn:p_spin_hamil}
\hat{H}_p(h) = -h \hat{S}_x - \frac{\Lambda}{p S^{p-1}}\hat{S}_z^p,
\end{equation}
where $h$ is the strength of the external magnetic field, $\Lambda$ the strength of the $p$-body interaction, and we have introduced the collective spin operators $\hat{\mbf{S}}=\sum_{i=1}^N \hat{\bm{\sigma}}^{(i)}/2$ with $\hat{\bm{\sigma}}$ a vector of Pauli operators. In terms of the collective spin operators, the symmetric subspace is spanned by the Dicke states, $\lvert S, M \rangle$, where $S=N/2$, resulting in a Hilbert space of dimension $N+1$, growing only linearly with the system size.

This family of models exhibits a rich and extensively studied variety of critical phenomena~\cite{Bapst2012}, including ground state quantum phase transition (GSQPT) between a paramagnetic and a ferromagnetic phase~\cite{Filippone2011}, excited state quantum phase transitions (ESQPT)~\footnote{For all the models with $p>2$, the phenomenology of the ESQPT is similar to that exhibited by the Lipkin model, see for example~\cite{Cejnar2021}} as well as dynamical quantum phase transitions (DQPT)~\cite{DelRe2016,Munoz-Arias2020,Correale2021}. The properties of these phenomena are markedly different depending on the value of $p$, and in the following we summarize the aspects which are of particular importance to our construction of FTCs. 

Critical phenomena in these models can be studied using a mean-field picture, which is exact in the thermodynamic limit $N\to\infty$ and describes the dynamics of the normalized mean spin $\mathbf{X} = \frac{\langle \hat{\mathbf{S}} \rangle}{S}$ on a unit sphere phase space. One tool to characterize this limit is given by the classical flow in phase space, $\dot{\mathbf{X}} = \mathcal{F}[\mathbf{X};h,\Lambda,p]$, which can be obtained via the Heisenberg equations motion for the spin operators after approximating $\langle A B\rangle\simeq \langle A\rangle \langle B\rangle$. Explicit expressions for this equations, which were previously studied in~\cite{Munoz-Arias2020}, are given in Appendix \ref{app:classical_equations}. A second complementary tool is given by the semiclassical energy function $E(\mathbf{X};h,\Lambda,p) = \frac{\langle H_p \rangle}{S}$, which serves the role of a free energy, where the average is taken over a spin coherent state $|\theta, \varphi\rangle = e^{-i\varphi \hat{S}_z}e^{-i\theta \hat{S}_y}|S, S\rangle$. For the $p$-spin models this energy function takes the form 
\begin{equation}
\label{eqn:semiclassical_energy}
E(\phi,Z;h,\Lambda,p) = -h\sqrt{1-Z^2}\cos(\phi) - \frac{\Lambda}{p}Z^p.     
\end{equation}

The critical points of different phase transitions correspond with certain structural changes of Eq.~(\ref{eqn:semiclassical_energy}) as a function of the ratio $\frac{h}{\Lambda}$. In the limit $\frac{h}{\Lambda}\to\infty$ the semiclassical energy is a single well with a minimum at $Z = 0$, indicating paramagnetic ordering. As the value of this ratio is reduced, Eq.~(\ref{eqn:semiclassical_energy}) might undergo a saddle-node bifurcation at which a saddle and a local minimum (metastable ferromagnetic phase) emerge. This point is typically referred to as the spinodal point~\cite{Goldenfeld2019}. Upon further reduction of the value of $\frac{h}{\Lambda}\to 0$, the local minimum reaches a point of degeneracy with the current global minimum, signaling the GSQPT critical point~\cite{Goldenfeld2019}, i.e., the point at which the ground state of the system changes character to ferromagnetic. The above scenario describes a first order GSQPT, as it is the case in the models with $p>2$~\cite{Bapst2012}. For the model with $p=2$, where one recovers the the Curie-Weiss paramagnet (a special case of the Lipkin-Meshkov-Glick or LMG model), the spinodal point and the GSQPT critical point coincide, and the transition from paramagnetic to ferromagnetic orderings is continuous or second order. For general $p$, the spinodal point can be shown to be (see Appendix~\ref{app:spinoal_and_criti})
\begin{equation}
\label{eqn:spino_point}
h W_{\rm spino}(p) = \Lambda, \enspace\text{with}\enspace W_{\rm spino}(p) = \sqrt{\frac{(p-1)^{p-1}}{(p-2)^{p-2}}},    
\end{equation}
and the GSQPT critical point is given by
\begin{equation}
\label{eqn:criti_point}
h W_{\rm GS}(p) = \Lambda, \enspace\text{with}\enspace W_{\rm GS}(p) = \frac{(p-1)^{p-1}}{\sqrt{(p(p-2))^{p-2}}}.
\end{equation}
The behavior of the spinodal and ground state critical points is illustrated in the $p$-spin phase diagrams of Fig. \ref{fig:spino_criti_system_sketch}a.

The emergence of a saddle point at the spinodal point implies the existence of a separatrix line in the classical phase space, which marks the boundary between two regions of distinct macroscopic motion of the mean spin. As a consequence of this, an extensive portion of the quantum spectrum is composed of states localized in the interior of the region enclosed by the separatrix line. In top-left panel of Fig. \ref{fig:reso_spheres} we show an example of the phase portrait for the $p=2$-spin. The dark lines indicate trajectories on the inside of the separatrix and the light lines indicate trajectories on the outside of it. The set of eigenstates which are localized in the regions inside the separatrix correspond to the extensive portion of the spectrum which have undergone a clustering ESQPT~\cite{Stransky2014,Stransky2015,Puebla2016,Cejnar2021}. These localized states will be used to define FTC phases in the present work, since they will lead to breaking the discrete time translation symmetry of the driven $p$-spin system. Thus, we will consider exclusively the regime of $h W_{\rm GS}(p) < \Lambda$, where the system is in the ferromagnetic phase. Finally, we point out that $p$-spin models have been studied in the context of so-called Boundary Time Crystals \cite{iemini2018,wang2021,piccitto2021}, in which the the continuous time-translation symmetry is broken by the presence of dissipation. In our work, we focus exclusively on studying Floquet Time Crystals in these models.

\subsection{Driving protocol and connection to kicked $p$-spin models}
\label{subsec:driving_protocol}
\begin{figure}[!t]
 \centering{\includegraphics[width=0.48\textwidth]{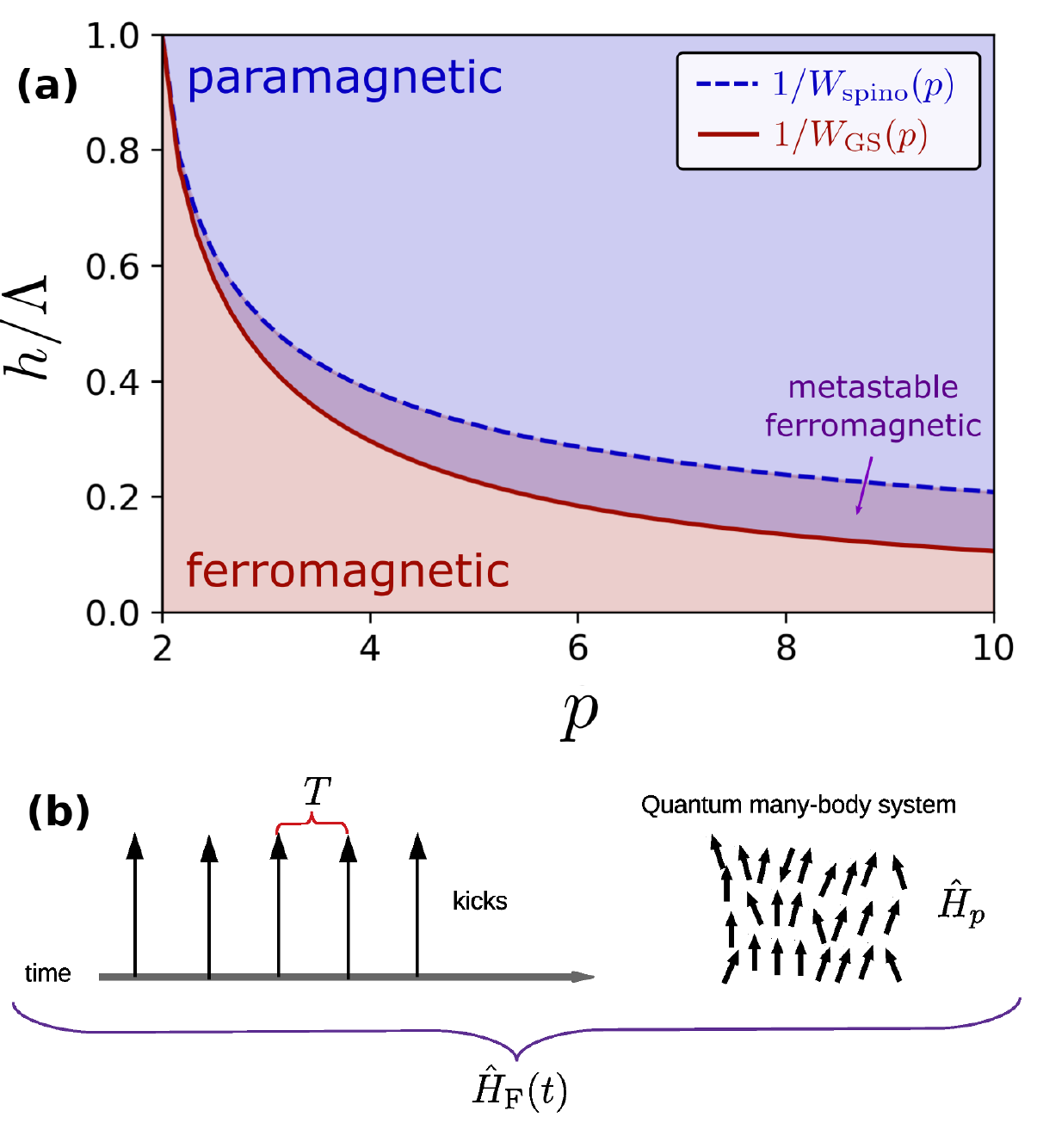}}
\caption{\textbf{(a)} Phase diagramas for the $p$-spin models described by Hamiltonian (\ref{eqn:p_spin_hamil}). We plot the spinodal, Eq. ~\ref{eqn:spino_point}, (blue dashed), and critical, Eq.~\ref{eqn:criti_point}, (solid red), lines for the $p$-spin family up to $p=10$, defining the paramagnetic (dark shaded area) and ferromagnetic (light shaded area) phases as a function of $h/\Lambda$ for the whole family. Notice the existence of a metastable ferromagnetic phase given by the area in between the two lines. \textbf{(b)} Sketch of the effect of the driving protocol in the clean many-body system. The time independent Hamiltonian gets mapped into a Floquet system with discrete time translation symmetry, \textit{i.e} $\hat{H}_F(t + T) = \hat{H}_F(t)$.}\label{fig:spino_criti_system_sketch}
\end{figure}
Given an initial state localized inside the region enclosed by the separatrix of a $p$-spin model, we consider an external periodic drive in the form of a train of short pulses or ``kicks'', as illustrated in Fig. \ref{fig:spino_criti_system_sketch}b. The period of this train of pulses is given by $T$, which we will fix to $T=1$, and at every period the system undergoes an instantaneous rotation around the $x$-axis by an angle $\alpha_{\rm B}$. A single evolution step is thus given by
\begin{equation}
    \label{eqn:floquet_ope}
    \hat{U}_{F} = e^{i\alpha_{\rm B} \hat{S}_x}e^{-i\hat{H}_{p}(h)},
\end{equation}
where we have taken $\hbar = 1$. From the point of view of the quantum (finite-size) system, the external drive transforms the Hamiltonian system into a Floquet system. Conversely, in the mean-field description, the evolution of the mean spin becomes stroboscopic and dictated by an area-preserving map (APM) ~\cite{Mackay1993} rather than a continuous phase space flow~\footnote{In fact, the mean spin evolves according to a ``coordinate transformation'' of the form $\mathbf{X}_{l+1} = \mathcal{A}[\mathbf{X}_l;\alpha_{\rm B}, h, \Lambda, p]$ with $l$ a discrete index accounting for the time step, the Jacobian being identical to $1$, hence the denomination of area preserving.}. This latter fact is of utter importance, as the mean-field description in terms of an APM brings in new phenomenology~\cite{Mackay1993} which is absent in the continous flow case. In particular, we will describe how the emergence of subharmonic system responses as a function of the drive parameters can be seen in phase space as the emergence of a resonance of the APM~\cite{Mackay1987}. A resonance is a region of phase space enclosed by separatrices connecting hyperbolic periodic points. Trajectories in the interior of the resonance correspond to macroscopic motion of the mean spin in the form of oscillations exhibiting a strong periodic subharmonic component. See Fig. \ref{fig:reso_spheres} for illustrations of this phenomenon. It is our aim to exploit the resonances of APMs in order to define FTC phases, and we will elaborate on this in the next section. 

Finally, let us point out an important connection between the Floquet operator in Eq.~(\ref{eqn:floquet_ope}) and the family of kicked $p$-spin models, recently introduced in Ref.~\cite{MunozArias2021}. In the case $h=0$, corresponding to a pure ferromagnetic $p$-spin, Eq.~(\ref{eqn:floquet_ope}) exactly recovers the kicked $p$-spin unitary. More generally, the connection can be made naturally as long as $h W_{\rm GS}(p) \ll \Lambda$. This is precisely the regime in which the initial states discussed in Sec. \ref{subsec:magnetic_model} are relevant. In this limit, we can rewrite the Floquet operator in Eq.~(\ref{eqn:floquet_ope}) as
\begin{equation}
    \label{eqn:kicked_p_spin_uni}
    \hat{U}_F = e^{i\alpha(h) \hat{S}_x}e^{i\frac{\Lambda}{pS^{p-1}}\hat{S}_z^p}, 
\end{equation}
which has the form of a kicked $p$-spin model. The equality holds up to order $\mathcal{O}(h\Lambda)$ provided $\Lambda$ is not too large, and we defined the modified precession angle $\alpha(h) = \alpha_{\rm B} + h$. The latter will allow us to examine the robustness of the phase, for fixed value of $\Lambda$, using $h$ as tunable parameter~\footnote{We allow us to consider both positive and negative values of $h$, provided $|h|W_{\rm GS} \ll \Lambda$ holds. This change in sign changes the character of the $p$-spin ground state from  ferromagnetic to anti-ferromagnetic, however the overall structure of the spectrum, \textit{i.e} the different ESQPT lines remain invariant, as this change in sign can be seen as the application of $e^{i\pi\hat{S}_y}$. Thus, we still have an extensive portion of excited states undergoing a clustering ESQPT and defining a class of states which is localized at the interior of the region enclosed by the separatrix.}. The APM for the mean-field limit of Eq.~(\ref{eqn:kicked_p_spin_uni}) has the form $\mathbf{X}_{l+1} = \mathcal{A}[\mathbf{X}_l;\alpha(h), \Lambda, p]$, and explicit expressions are given in  Appendix 
~\ref{app:classical_equations} (see also~\cite{MunozArias2021} for details).

\section{Resonances, resonance conditions and bifurcations in area preserving maps}
\label{sec:reso_reso_cond_bifus}
In this section, we develop a framework to predict and describe FTCs in mean field models. First, we discuss so-called resonances in APMs, which are the basic structure leading to subharmonic periodic response in the system. We derive conditions for the existence of these resonances by analyzing the time-dependent periodic Hamiltonian underlying the Floquet map. From this, we show how a new symmetry emerges in the driven system at these resonance conditions. Finally, we argue that not all resonances will lead to the extensive structural changes in phase space needed to define a robust FTC, and propose that only resonances that can be linked to bifurcations (i.e. resonant bifurcations) in the APM will lead to proper FTC behavior.

\subsection{Resonances in area preserving maps}
\label{subsec:resonances_in_apm}
\begin{figure*}[!t]
\centering{\includegraphics[width=0.95\textwidth]{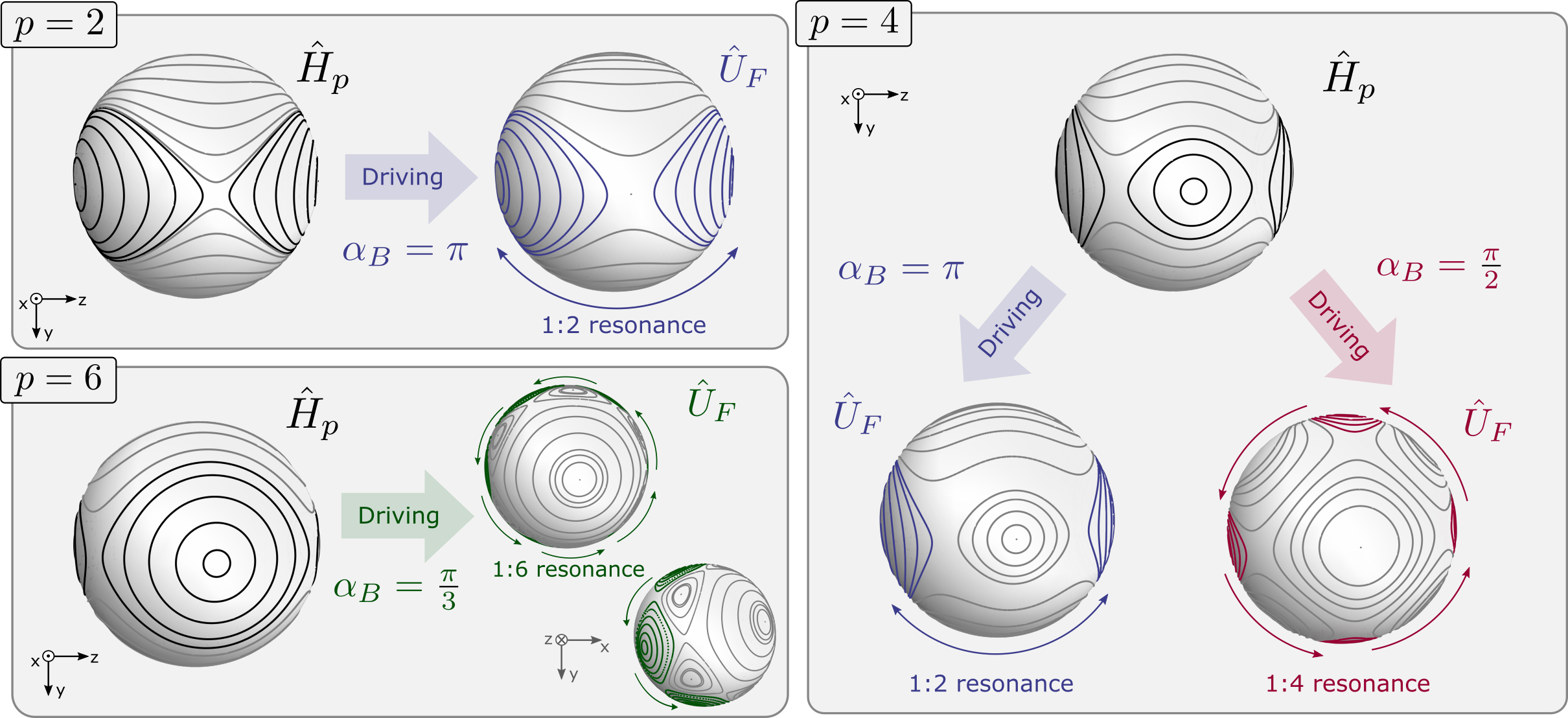}}\caption{Trajectories of the phase space flow of the $p$-spin model and the area preserving map of the driven $p$-spin model (kicked $p$-spin), for three different values of $p$: $p=2$ \textbf{(top-left)}, $p=4$ \textbf{(right)}, and $p=6$ \textbf{(bottom-left)}, and their respective values of the angle of the drive $\alpha_{\rm B}$. In each panel the left (top) sphere corresponds to the trajectories of the flow, dark lines highlight the trajectories at the interior of the separatrix, light lines are in the exterior. We explicitly show the effects of the action of the drive in mapping the flow into an area preserving map with the corresponding resonances, trajectories belonging to the $1$:$2$, $1$:$4$, and $1$:$6$ are shown in blue, red and green, respectively.}
\label{fig:reso_spheres}
\end{figure*}

A resonance of an APM is a region of finite area in phase space delimited by separatrices connecting hyperbolic (periodic) points. Every resonance has a central periodic orbit with period equal to that of the resonance or some integer multiple of it~\cite{Mackay1987}. Therefore, the trajectories in the interior of the resonance exhibit a strong subharmonic behavior with period $q$, with $q>1$ and integer. Consider a $1$:$q$ resonance, for an initial condition on or near the separatrix. Then, the stable or unstable character of the separatrix branches will only be apparent after $q$-steps of the stroboscopic evolution. Furthermore any initial condition inside the resonance will only look localized, as the parent trajectories of the classical flow of the Hamiltonian system, every $q$-steps of the stroboscopic evolution. For instance, take the system with $p=2$, the phase portrait for $\hat{H}_p$ is shown in upper-left panel of Fig.~\ref{fig:reso_spheres} (left sphere). In the undriven case, the motion of a state initially localized inside one of the lobes remains confined to that lobe. Once driving is included such that a resonance condition is met, the same type of initial state will now  visit all other lobes before coming back to the initial one. In this case, $\alpha_{\rm B} = \pi$ and this process happens after $q=2$ steps. Similar scenarios occur for other values of $q$, see Fig.~\ref{fig:reso_spheres}.

From this description of the classical map, we see that the macroscopic motion of our chosen set of initial states will be dominated by the central periodic orbit, and hence generic physical observables will display discrete time translation symmetry breaking. The existence of a resonance of the APM implies that the Floquet states, which in the limit $S\to\infty$ will correspond to trajectories inside the resonance, must have support on all lobes of the resonance, and hence could be written as superpositions of macroscopically distinct states, each of which has support inside \textit{one} of the lobes of the resonance. Thus, a subset of Floquet states will look like "cat-states'', and thus display eigenstate ordering in phase space, a key requirement needed for the definition of an FTC phase~\cite{Else2020}~\footnote{If one finds Floquet states inside the resonance whose support is not in all lobes of the resonance, they will have a nontrivial evolution every step and thus will not satisfy an eigenvalue equation for the Floquet operator.}.

Using the resonances of an APM to define FTC phases brings in the robustness of the phase for free. That is, the observed subharmonic response will be robust to small changes in the system parameters, a statement that follows from the persistence of the fixed points. The latter concept can be explained as follows: Let us consider the APM $\mathbf{X}_{l+1} = \mathcal{A}[\mathbf{X}_l;\alpha(h), \Lambda, p]$ giving the evolution of the mean spin in the thermodynamic limit. The periodic orbits on a resonance, hyperbolic or elliptic, are fixed points of $\mathcal{A}^q[\mathbf{X}_l;\alpha(h), \Lambda, p]$ with $q$ the period of the resonance. A result in the theory of APMs guarantees the persistence of fixed points~\cite{Mackay1993}, that is, the resonance and its periodic orbits persist under the action of small perturbations, at most being moved to some new location in phase space, unless the tangent map evaluated at the fixed point has an eigenvalue equal to $1$. If the latter condition holds, the map undergoes a bifurcation and the periodic orbits might disappear. 

It is desirable to have conditions on the system parameters for which emergence of a resonance is guaranteed. In the following we discuss two of these situations: resonance conditions and bifurcations in APMs.

\subsection{Resonance conditions}
\label{subsec:resonance_conditions}
Given an APM which describes the dynamics of the system, we can always assign a time-periodic Hamiltonian, $\hat{H}(t) = a\hat{H}_1 + b\hat{H}_2\sum_{n=-\infty}^\infty \delta(t-nT)$, giving rise to the same APM equations in the thermodynamic limit. Then, a resonance condition is given by the values of $a,b$ for which the system receives an integer number of kicks in the form of $\hat{H}_2$ during the time it takes to complete a full cycle of the dynamics generated by $\hat{H}_1$.

For the Floquet system in Eq.~(\ref{eqn:floquet_ope}) a resonance condition is given by the value of the angle of the drive for which the mean spin receives an integer number of kicks during the time it takes to complete a full period of the trajectory associated with the corresponding phase space flow. Correspondingly, for the kicked $p$-spin in Eq.~(\ref{eqn:kicked_p_spin_uni}), it is given by the situation at which the mean spin receives an integer number of kicks during the time it takes to complete a single precession, that is, when $\alpha_{\rm B} = \frac{2\pi}{q}$ with $q$ an integer. At a resonance condition, what used to be separatrices emerging from saddle points of the associated continuous flow, and enclosing regions of phase space where eigenstates are localized~\footnote{In the case of even $p$-spin models this localized eigenstates are symmetry broken states, breaking the $\mathbb{Z}_2$ symmetry of the Hamiltonian.}, now become separatrices connecting hyperbolic periodic points. The region enclosed by the separatrices becomes a $1$:$q$ resonance of the APM~\cite{Mackay1987}.

To explore the consequences of a resonance condition in our driven system we define the time-dependent Hamiltonian
\begin{equation}
\label{eqn:kicked_p_spin_hamil}
\hat{H}(t) = -\alpha_{\rm B} \hat{S}_x - \frac{\Lambda}{pS^{p-1}}\hat{S}_z^p \sum_{n = -\infty}^\infty \delta (t - n),
\end{equation}
where we have fixed the period of the drive to be $T = 1$. Notice that this corresponds to the kicked $p$-spin Hamiltonian~\cite{MunozArias2021}. Eq.~(\ref{eqn:kicked_p_spin_hamil}) can be brought into the form $\hat{H}(t) = \hat{H}_{\rm reso} + \hat{V}(t)$ (see Appendix \ref{app:resonant_hami_symmetries} for details), where $\hat{H}_{\rm reso}$ is the resonance Hamiltonian and $\hat{V}(t)$ a time dependent perturbation which is typically of high frequency and whose effect vanishes on average.

The resonance Hamiltonian encodes the effects that the resonance condition induces in the Floquet system. Its form its derived in Appendix \ref{app:resonant_hami_symmetries}, and it reads 
\begin{equation}
\label{eqn:resonance_hamiltonian}
\hat{H}_{\rm reso} = -\frac{\Lambda}{q p S^{p-1}} \sum_{j = 1}^{q} \left( \hat{\mathbf{O}}_{YZ} \cdot \vec{e}_j\right)^p,
\end{equation}
where $\hat{\mathbf{O}}_{YZ} = (\hat{S}_y, \hat{S}_z)$ is the projection of the collective spin onto the $y$-$z$ plane, and $\vec{e}_j = \left(-\sin\left(\frac{2\pi}{q}j \right), \cos\left(\frac{2\pi}{q}j \right) \right)$ are the vertices of a $q$-regular polygon (see for instance~\cite{Chernikov1989} and chapter 6 of~\cite{Zaslavsky1991}). The resonance Hamiltonian is a sum of $q$ terms in the form of a $p$-twist each along one of the directions $\vec{e}_j$. The form of $\hat{H}_{\mathrm{reso}}$ implies that a region of finite size in the vicinity of the great circle in the $y$-$z$ plane develops a $1$:$q$ resonance, whose central periodic orbit has its points on the vertices of a $q$-regular polygon. This resonance then satisfies an emergent $q$-fold, $\mathbb{Z}_q$, rotational symmetry, which can be appreciated in the phase space portraits of Fig. \ref{fig:reso_spheres}~\footnote{In fact the resonance has a symmetry dictated by the dihedral group $\mathrm{D}_q$. Extensive work investigating emergent symmetries at resonance conditions in some area preserving maps \textit{e.g} delta-kicked harmonic oscillator, and their associated phase space structures, including crystalline and quasi-crystalline lattices, has been done by Zaslavsky, Sagdeev, Usikov and Chernikov~\cite{Chernikov1989,Zaslavsky1991}.}.

One could, then, use every resonance condition of Eq.~(\ref{eqn:kicked_p_spin_uni}) to define an FTC phase, since an appropriate choice of the initial state and observable will reveal the subharmonic character of the system response, which will be robust to small changes in $h$ (leading to small changes in $\alpha_{\rm B}$ away from $\frac{2\pi}{q}$, as discussed in Sec. \ref{subsec:resonances_in_apm}). This is in fact the method employed in~\cite{nurwantoro2019} and~\cite{pizzi2021} for the identification of higher period FTC phases, which we have put into a more formal footing. However, the emergent symmetry discussed above is a feature of only a region of phase space in the vicinity of the $y-z$ plane. Following previous works~\cite{Else2020}, we require the emergent symmetry to be global in order to define a proper FTC. For mean-field models, this implies that the new symmetry affects the entirety of phase space (or a majority fraction of it). It is desirable, then, to introduce further requirements on the system parameters in order to position a given resonance condition as a point where a proper FTC phase exists. To this end we investigate the dynamics near the poles, \textit{i.e} around the points $X = \pm1$ and their bifurcations~\footnote{Notice that we do not need to analyze the bifurcations of a higher iteration of the map, as those higher period bifurcations will be seen, in the bare map, as points for which the eigenvalues of the tangent map are some root of unity, see for instance~\cite{Mackay1993,Meyer1970}.}.

\subsection{Bifurcations} 
\label{subsec:bifurcations}
For an area preserving map, $1$:$q$ resonances can also emerge as consequences of a bifurcation process of a fixed point of the map. These, generally, are of the types described in the classification of generic bifurcations~\cite{Meyer1970,Mackay1993}. Given a fixed point of the map, say $\mathbf{X}_{\rm fix}$, if the eigenvalues of $\left. \frac{\partial \mathcal{A}}{\partial\mathbf{X}_l}\right|_{\mathbf{X}_{\rm fix}}$ are equal to $1$, $-1$, or the $q$th root of unity, then the fixed point undergoes a tangent, period doubling, or period $q$ bifurcation, respectively. In the case of period doubling or period $q$ bifurcation, one also observes the emergence of a $1$:$q$, $q\ge2$, resonance in phase space.

In~\cite{MunozArias2021} bifurcations of the fixed points at the poles $X = \pm1$ were analyzed. It was shown that the kicked system with $p=2$ has a period doubling bifurcation at $\alpha_{\rm B} = \pi$, and all the other models with $p>2$ have $d$-$q$ bifurcations whenever $\alpha_{\rm B} = \frac{2\pi d}{q}$ with $d,q$ relative primes, $q\ge2$ and $q>d$ (expressions for the tangent map eigenvalues are given in Appendix \ref{app:classical_equations}). However, not all the allowed bifurcations in the classical limit introduced large structural changes with signatures in the quantum system. This can be understood by analysing the structure of the multi-body interaction term $\sim \hat{S}_z^p$. In terms of raising and lowering operators $\hat{S}_\pm = \hat{S}_y \pm i\hat{S}_z$, the interaction term will connect states in the $\hat{S}_x$ basis which are $l$-flips away, where $l \in \left[p, p-2, p-4, ... , 2(1)\right]$, and the set ends at $2 (1)$ for even (odd) $p$-spins, that is, in the basis of $\hat{S}_x$ it introduces a coupling between states whose spin projection differ by $l$-units. Therefore, from all the discrete rotational symmetries emerging at resonance conditions of the term $\sim\alpha(h)\hat{S}_x$, only few of them are permitted by the interaction, which for the case of $p>2$ are also bifurcation points of the poles. From this considerations, we can define a reduced set of bifurcations
\begin{equation}
\label{eqn:set_of_bifus}
\mathcal{B}^{(p)}_{\rm bifu} = \left\{ \frac{2\pi}{m} : m\in\left[p, p-2, p-4, ..., 2(1)\right] \right\}.
\end{equation}
We refer to the resonance conditions which are also bifurcation points and are contained in $\mathcal{B}_{\rm bifu}^{(p)}$ as resonant-bifurcations. This set was previously identified in~\cite{Chinni2021}, where, using unitary perturbation theory, it was shown that large structural changes take place around these points \footnote{Importantly, for the emergence of FTC phases one should also account for bifurcation points having multiplicities larger than one. If $\frac{2\pi}{m} \in \mathcal{B}^{(p)}_{\rm bifu}$, and $r\frac{2\pi}{m} < \pi$ for $r\in\mathbb{N}$, then that bifurcation point should also be included in $\mathcal{B}^{(p)}_{\rm bifu}$. In the case of even $p$-spins this is a consequence of the inherited $\mathbb{Z}_2$ symmetry, since it forces all odd period bifurcations to be double, thus they will essentially look as a even period bifurcation with twice the period, see~\cite{MunozArias2021} for details. For instance, when $p=6$ we have $\mathcal{B}^{(6)}_{\rm bifu} = \{ \frac{2\pi}{6}, \frac{2\pi}{4}, \frac{2\pi}{2}\}$, however $ 2\times\frac{2\pi}{6} = \frac{2\pi}{3} < \pi$, we should include $ \frac{2\pi}{3}$ in $\mathcal{B}^{(6)}_{\rm bifu}$.}.

At resonant-bifurcation points, not only a finite region in the vicinity of the $y$-$z$ plane exhibits an emergent $q$-fold rotational symmetry, but also these regions extend all the way up to the poles, making it a global symmetry of phase space. When the resonance condition is no longer satisfied, $\alpha_{\rm B} \notin \mathcal{B}_{\rm bifu}^{(p)}$, the linear term $\sim h \hat{J}_x$ introduces a tilt which displaces the resonance on the $y$-$z$ plane to a new location, and the time-independent part of the Hamiltonian now reads $h\hat{S}_x + \hat{H}_{\rm reso}$. Under these conditions, a wider region of phase space contains trajectories with a strong period-$q$ component. These come from both resonances: those coming from the resonance condition and those emerging as a consequence of the bifurcation, and from all the trajectories in between them. Even though not all phase space exhibits a $q$-fold rotational symmetry, a large portion of it does, and so it is an approximate symmetry. This approximate symmetry provides ground for the definition of an FTC phase.

\subsection{Classical picture of Floquet time crystal phases in $p$-spin models}
\label{sec:class_pic_ftc}
Let us now collect all the different ideas presented in the previous sections into a single summary of a classical picture of Floquet time crystals in $p$-spin models.

Consider an initial state supported inside the ferromagnetic region of phase space, \textit{i.e} the region enclosed by the classical separatrix of the phase space flow, of a $p$-spin model. When the system is driven, this region is transformed into a resonance of the APM, where now separatrices connect hyperbolic periodic points. The ensuing dynamics of the initial state is no longer confined inside the original support, but rather visits all the different lobes of the resonance. In fact, the system will undergo oscillations with the period equal to that of the central orbit of the resonance, returning to the region of the original support every $q$ applications of the drive, with $q$ the period of the resonance. Thus, the period of the resonance dictates the type of FTC behavior emerging as a consequence of the drive.

We have argued, and will provide evidence, that if the system has $p$-body interactions, with $p$ large enough, then by tuning the angle of the external drive to one of the values in the set $\mathcal{B}_{\rm bifu}^{(p)}$, FTC phases beyond a period double $2T$-FTC are accessible. These special angles are resonance conditions at which an emergent dynamical $q$-fold rotational symmetry arises in the system, leading to a special reconfiguration of Floquet states and Floquet phases, as we will discuss in next section. They are also bifurcation points, therefore the resonances which exist in phase space can only disappear via a second bifurcation process, providing a certain degree of robustness to the emerging FTC phases. The resonant-bifurcation points are then the key element we exploit in order to define $qT$-FTC phases, where $0<q\le p$ as discussed above. In the remaining of the manuscript we will characterize, both for a finite system sizes and in the thermodynamic limit, the different $qT$-FTC phases emerging in periodically driven $p$-spin models.

\section{Characterizing Floquet Time Crystal Phases}
\label{sec:characterization}
In this section we discuss the different diagnostic tools we will use in later sections to characterize FTC behavior in driven $p$-spin models. An important figure of merit will be the presence of subharmonic response of $f_{Z}(t) = \langle\psi(t)|\frac{\hat{S}_z}{S}|\psi(t) \rangle$, identified via its power spectrum. In addition, we will study the structural changes in the Floquet operator associated with the onset of eigenstate ordering via analyzing its spectral statistics. Furthermore, this analysis will allow us to establish a connection between the FTC transition and a clustering excited state quantum phase transition in the mean-field limit. Then, we propose the use of out of time order correlators (OTOCs) as an order parameter to detect the transition between an FTC and non-FTC, and connect this to the theory of dynamical quantum phase transitions. Finally, owing to the mean-field nature of these models, we discuss how to characterize the FTC phase diagrams in the thermodynamic limit by studying the phase-space average of the dynamical response of the system.

\subsection{Eigenstate ordering, emergent dynamical symmetries and spectral statistics}
\label{subsec:eigenstate_ordering}
\begin{figure}[!t]
 \centering{\includegraphics[width=0.46\textwidth]{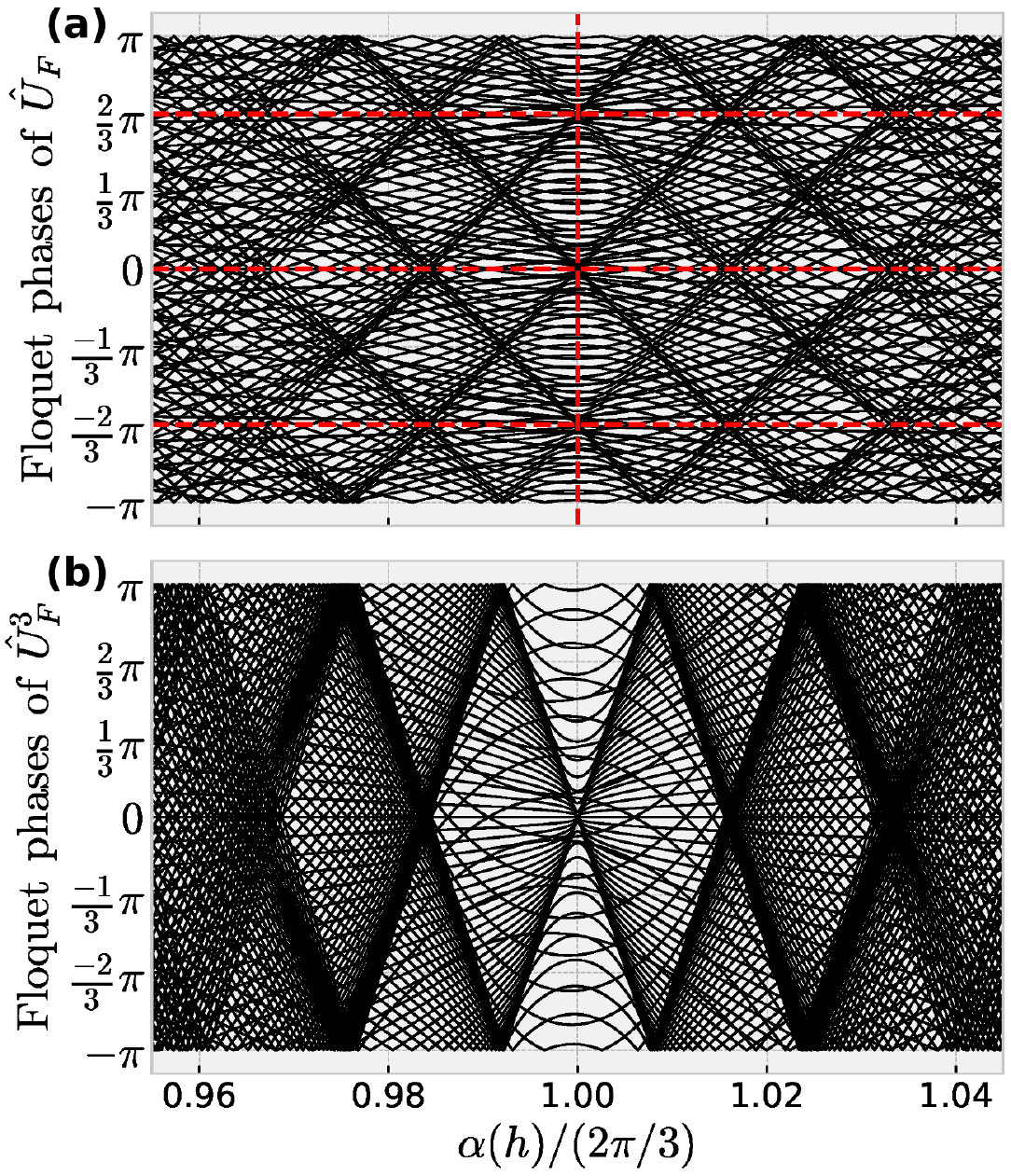}}
\caption{Example of level clustering inside an FTC phase using the system with $p=3$. \textbf{(a)} Floquet phases of $\hat{U}_F$ in the vicinity of $\alpha(h) = \frac{2\pi}{3}$, the horizontal lines show the values $\frac{2\pi}{3}, 0, -\frac{2\pi}{3}$, notice the symmetry of the spectrum with respect to this lines. The reorganization of the spectrum into a regular structure indicates the emergence of a symmetry and $\frac{2\pi}{3}$-pairing of Floquet phases. \textbf{(b)} Floquet phases of $\hat{U}_F^3$. The $\frac{2\pi}{3}$-pairing is now evident in the central section of the spectrum. Paramaters are: $p=3$, $\Lambda = 0.5$, $N = 128$.}
\label{fig:example_level_cluster}
\end{figure}
Eigenstate ordering of the Floquet states refers to a special structure of the eigenvectors of the Floquet operator $\hat{U}_F$, which have the form of cat-like states, \textit{i.e} superpositions of macroscopically distinct states~\cite{else2016,Else2020}. In the mean-field limit, this translates to Floquet states which have support in all of the $q$ lobes of a $1$:$q$ resonance, that is, superpositions of $q$ states each of which is localized in one of the lobes of the resonance. Therefore a portion of the Floquet spectrum is composed of groups of $q$ Floquet phases, where each of the Floquet phases in one of these groups define the vertices of a $q$-regular polygon in the unit circle. 

In the case of $p=2$ and $\alpha(h)\simeq \pi$, \textit{i.e} the FTC phase in the LMG model~\cite{russomanno2017}, this phenomenon was referred to as $\pi$-pairing of Floquet phases. More precisely, the Floquet states with support inside the $1$:$2$ resonance come in pairs, with each member of a pair having a definite parity under $e^{i\pi \hat{S}_x}$. The eigenphases of Floquet states in one of these pairs are located at diametrically opposed points on the unit circle and thus they differ by $\pi$. As a consequence, when considering the operator $\hat{U}_F^2$ the Floquet phases corresponding to Floquet states in those pairs will be degenerate. Thus, the spectral statistics of $\hat{U}_F^2$ will deviate from the expected Poissonian behavior~\footnote{Here Poisson statistics of the adjacent level ratios is expected as we are in a parameter regime where classical system is regular.} due to a strong degeneracy arising from to the clustering of levels by pairs.

More generally in FTC phases of period $q$ with $q>2$ we expect to see $\frac{2\pi}{q}$-pairing of the Floquet phases of $\hat{U}_F$ as the manifestation of eigenstate ordering. This pairing will lead to a strong degeneracy in the Floquet phases of $\hat{U}_F^q$, with its spectral statistics deviating from that of a Poisson distribution. In Fig. \ref{fig:example_level_cluster} we show an example of the Floquet phases for the system with $p = 3$ around the region of the $3T$-FTC phase. Notice the emergent apparent regularity of the spectrum in Fig. \ref{fig:example_level_cluster}a, and how the $\frac{2\pi}{3}$-pairing is revealed through the degeneracy of the Floquet phases of $\hat{U}^3_F$ in Fig. \ref{fig:example_level_cluster}b. The existence of this degeneracy is a signature of an emergent dynamical symmetry, a discrete $q$-fold rotational symmetry, which is also manifested in the form of the effective Hamiltonian of $\hat{U}_F^q$, give by $h\hat{S}_x + \hat{H}_{\rm reso}$ as discussed in Sec.~\ref{subsec:resonance_conditions}. To describe this behavior we will use the average adjacent spacing ratio~\cite{Atas2013}, a standard measure of spectral statistics, defined by 
\begin{equation}
 \label{eqn:adjacent_ratio}
 \overline{r} = \frac{1}{N+1}\sum_{j=1}^{N+1}r_j, \quad r_j = 
\frac{{\rm min}(d_j, d_{j+1})}{{\rm max}(d_j, d_{j+1})},
\end{equation}
where $d_j = \mu_{j+1} - \mu_j$ is the eigenphase spacing and $N = 2S$. Further, we define the normalized average adjacent spacing ratio $\tilde{r} = \overline{r}/r_{\rm POS}$ where $r_{\rm POS}$ is the value for a Poisson distribution, hence any value of $\tilde{r} < 1$ will indicate a high degree of eigenphase clustering, signaling the onset of the FTC phase.

As stated above, the emergence of $qT$-FTC can be diagnosed by the clustering of an extensive portion of the eigenspectrum of $U_F^q$. This phenomenon admits an interesting connection to a clustering excited-state quantum phase transition (ESQPT)~\cite{Cejnar2021}. For a time-independent Hamiltonian with a one degree of freedom, a clustering ESQPT can be identified by a series of vanishing energy gaps in the spectrum, forming a continuum in the thermodynamic limit. Due to the closing gaps, the energy levels cluster around these regions, and the density of states diverges~\cite{Stransky2014,Stransky2015,Cejnar2021}. ESQPTs have been mostly identified in systems with clear-cut classical limits (like the $p$-spin models), and their existence is intimately connected to the emergence of separatrices in the mean-field phase space. Even though the $p$-spin models present ESQPTs, the phenomena shown here in the \textit{driven} $p$-spin models is different, since the associated classical phase spaces are different. In general, the emergence of ESQPTs in driven quantum systems is a relatively unexplored phenomenon, with only few examples~\cite{Bastidas2004,Bandyopadhyay2015,GarciaMata2021}. 

From a pure kinematic point of view, the nonFTC-FTC transition, as seen by the emergence of eigenstate ordering, displays all the features of a clustering ESQPT~\cite{Stransky2015}. For a system whose thermodynamic limit is described by a single degree of freedom, a clustering ESQPT corresponds to a singularity of the density of states in the form of a logarithmic divergence. This divergence originates in the diverging oscillation periods of trajectories corresponding to initial conditions which get closer and closer to a separatrix line as a function of a control parameter. Therefore, when one examines the energy spectrum, it is crossed by a curve along which excited states cluster together towards degeneracy. For our system this observation follows naturally from the mean-field phase space analysis made in Sec. \ref{sec:reso_reso_cond_bifus}, as the nonFTC-FTC transition, in this limit, is accommodated by the emergence of resonances of the APM. Let us assume a $1$:$q$ resonance leading to a $qT$-FTC, recall that under the effective representation of the dynamics every $q$ steps the partial separatrices connecting hyperbolic periodic points behave as actual separatrices emerging from saddle points. In fact, they correspond to separatrices of the effective Hamiltonian of $\hat{U}_F^q$, given by $\hat{H}_{\rm eff}(h) = h\hat{S}_x + \hat{H}_{\rm reso}$, with $h$ playing the role of control parameter. As such the density of states will display a logarithmic divergence, a phenomenon we discuss in more detail in Appendix ~\ref{app:kicked_dos}. 

\subsection{Out of time order correlators as dynamical order parameters for the nonFTC - FTC transition}
The structural change of an extensive portion of the Floquet states from lacking eigenstate ordering to displaying eigenstate ordering has a direct consequence in the dynamics. In fact, the emergence of an extensive $\frac{2\pi}{q}$-pairing of the Floquet phases makes $\omega = \frac{2\pi}{q}$ to be the dominant frequency in the system response. As such, the macroscopic motion of the mean spin undergoes a drastic change, going from approximating Larmor precession at some frequency, whose value is in principle arbitrary and changes continuously with the system parameters, to be precessing with a frequency locked at the $q$th subharmonic of the frequency of the drive. This change in the macroscopic motion can be thought of as a dynamical quantum phase transition (DQPT)~\cite{Sciolla2011,Zunkovic2016,Heyl2018rev}, and technically its associated with an order parameter DQPT, or DQPT of type-I~\cite{Zunkovic2018}. We notice that, the drastic change in the macroscopic motion concerns only the frequency of precession and not the precession axis, therefore the asymptotic value of the quasi-steady state of an observable which begins out of equilibrium remains unchanged, having the consequence that the ``usual" DQPT order parameters, such as the time-averaged magnetization and the time-averaged two point correlation function, fail at detecting this dynamical transition.

In order to have a faithful identification of this DQPT we resort to higher-order correlation functions, in particular out of time order correlators (OTOCs). OTOCs are widely used in studies of out-of-equilibrium dynamics and information scrambling~\cite{Swingle2018,Riddell2019,Pappalardi2018}. Recently, it has been shown that in cases where a dynamical order parameter might not be well defined or it is difficult to construct, the long-time average of the OTOC can be used to dynamically detect quantum phase transitions, be it of equilibrium or dynamical nature~\cite{Heyl2018,BurkaDag2019,Wang2019}. The OTOC is defined as 
\begin{equation}
\label{eqn:def_otoc}
\mathrm{F}_{W,V}(t) = \langle \hat{W}^\dagger(t) \hat{V}^\dagger \hat{W}(t) \hat{V} \rangle
\end{equation}
where $\hat{W}(0)$ and $\hat{V}$ are two operators which commute and $\hat{W}(t=lT) = \hat{U}_F^{\dagger l}\hat{W}(0)\hat{U}_{F}^l$ is the Heisenberg evolution of $\hat{W}(0)$, and the average is taken on some reference initial state. We consider the infinite time average of Eq. (\ref{eqn:def_otoc}): 
\begin{equation}
\label{eqn:long:otoc}
\mathrm{F}_{W,V}^\infty = \lim_{t\to\infty}\frac{1}{t} \int_0^{t}\mathrm{F}_{W,V}(t')dt'.
\end{equation}
We will be focusing on OTOCs with operators $\hat{W}(0) = \hat{V} = \frac{\hat{S}_z}{S}$, which we denote $\mathrm{F}_{Z,Z}(t)$ and $\mathrm{F}_{Z,Z}^\infty$, where the average is taken over an ``infinite temperature" state $\rho_0 = \frac{\mathbb{I}}{N+1}$. Using the long-time average of the OTOC as a diagnostic, we will see that the transition between non-FTC to FTC behavior can be regarded as a DQPT, where $\mathrm{F}_{Z,Z}^\infty = 0$ in the non-FTC phase and $\mathrm{F}_{Z,Z}^\infty \ne 0$ inside the FTC phase.

\subsection{Averaged classical response and mean-field phase diagram}
\label{subsec:measure_for_classical_phase_diagram}
To complement the characterization of the FTC phases made via the spectral statistics and the OTOC, we study phase diagrams of our FTC phases in the thermodynamic limit. These will be constructed based on the dominant frequency of the mean response of the system, obtained via the phase space average of the long time evolution of the classical $ZZ$ correlation function. This correlation function is defined as 
\begin{equation}
\label{eqn:avg_correlation}
\mathcal{C}_{\rm PS}^Z(l) = \left\langle \lim_{l\to\infty}  \mathcal{C}_l^Z  \right\rangle_{\rm PS} = \left\langle\lim_{l\to\infty} Z_{l}Z_0 \right\rangle_{\rm PS},
\end{equation}
where $Z_l$ is the stroboscopic evolution of the $z$-coordinate of the mean spin, and the brackets $\langle.\rangle_{\rm PS}$ indicate phase space average. Furthermore, to directly observe the subharmonic character of the system response we investigate the power spectrum $|\mathcal{C}^Z_\omega|^2$ of the averaged correlation function, where $\mathcal{C}^Z_\omega$ is given by 
\begin{equation}
\label{eqn:pow_avg_correlation}
\mathcal{C}_\omega^Z = \sum_{l = 0}^{T_{\rm max}} \mathcal{C}^Z_{\rm PS}(l) e^{-i\omega l},
\end{equation}
the discrete Fourier transform of the finite-time averaged correlation performed over $T_{\rm max} \gg 1$ periods. A phase diagram as a function of $\alpha(h)$ and $\Lambda$ can then be constructed by identifying the regions in parameter space where $|\mathcal{C}^Z_\omega|^2$ presents a peak at $\omega=2\pi/q$. The actual quantity to be computed reads: 
\begin{equation}
\label{eqn:class_phase_diagram}
\mathcal{G}(\Lambda, \alpha(h)) = \begin{cases}
\frac{|\mathcal{C}_\omega^Z|^2}{\max|\mathcal{C}_\omega^Z|^2} & \text{if}\enspace \underset{\omega}{\arg\max}|\mathcal{C}_\omega^Z|^2 = \frac{2\pi}{q}, \\
\enspace\quad 0 & \text{otherwise},
\end{cases}
\end{equation}
where the normalization is included for convenience, since for systems with coexisting FTC phases, the $2T$-FTC might overshadow all the other phases. 

As mentioned in the previous subsection, the change in the macroscopic motion only concerns the frequency of oscillation, as such the asymptotic value of classical two point correlation functions in the dynamical steady state, remains unchanged. However, we are constructing phase diagram based on the frequency of this correlation. Furthermore, the phase space average will, in absence of a single strong global frequency, acts as a classical dephasing process. However, inside an FTC we expect a large portion of phase space to be populated by trajectories whose dominant frequency is the $q$th subharmonic of the drive frequency. Therefore, the existence of an non-negligible peak at $\omega = \frac{2\pi}{q}$ in the phase-space-averaged system response provides strong evidence of the extensivity of the FTC under study.

\begin{figure*}[!ht]
 \centering{\includegraphics[width=0.98\textwidth]{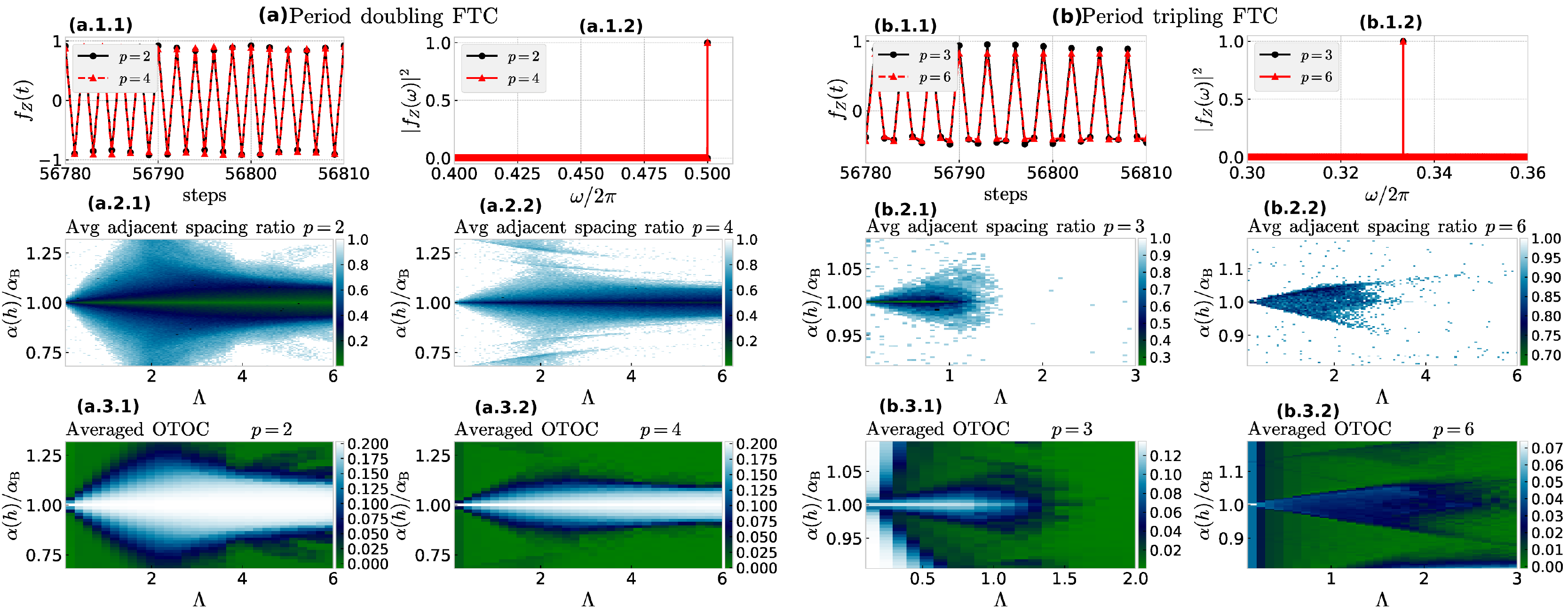}}
\caption{Characterization of the $2T$-FTC \textbf{(a)}, and $3T$-FTC \textbf{(b)} phases in the family of models $p$-spin models. \textbf{(a.1) - (b.1)} Snapshot of $f_Z(t)$ in the long time regime and its normalized power spectrum. A clear subharmonic response with $\omega/2\pi = \frac{1}{2}, \frac{1}{3}$ is observed in \textbf{(a)} and \textbf{(b)}, respectively. \textbf{(a.2) - (b.2)} $\tilde{r}$ as a function of $\alpha(h)$ and $\Lambda$ in the vicinity of $\alpha_{\rm B} = \pi$ and $\alpha_{\rm B} = \frac{2\pi}{3}$, respectively. White indicates Poisson statistics, nonwhite indicates some degree of pair/triplets clustering of levels and indicates the region of parameters where the system behaves as a $2T$-FTC and $3T$-FTC, respectively. \textbf{(a.3)-(b-3)} $\mathrm{F}^\infty_{Z,Z}$ as a function of $\alpha(h)$ and $\Lambda$ in the vicinity of $\alpha_{\rm B} = \pi$ and $\alpha_{\rm B} = \frac{2\pi}{3}$, respectively. Green indicates $\mathrm{F}^\infty_{Z,Z} = 0$, nongreen indicates $\mathrm{F}^\infty_{Z,Z} \ne 0$, hence the region of parameters where the system behaves as a $2T$-FTC and a $3T$-FTC, respectively. Parameters are: $N = 1024$ in \textbf{(a.2)} and \textbf{(b.2)}, and $N = 256$, $T_{\rm max} = 16000$ in \textbf{(a.3)} and \textbf{(b.3)}. $h = 0.1$, $\Lambda = 0.7$, $T_{\rm max} = 60000$, $N = 1024$ in \textbf{(a.1)} and $h = 0.05$, $\Lambda = 0.7$, $T_{\rm max} = 60000$, $N = 1024$ in \textbf{(b.1)}.}
\label{fig:peroid_doubling_tripling}
\end{figure*}

\section{Emergence of Floquet time crystal phases around bifurcation points}
\label{sec:emergence_ftc}
In this section we present and discuss the numerical results which characterize the emergence of various FTC phases in driven $p$-spin models. As discussed previously, these will appear in the vicinity of the points in $\mathcal{B}_{\rm bifu}^{(p)}$, the set defined in Eq. (\ref{eqn:set_of_bifus}). We will present results on the $1$-$2$, $1$-$3$ and $1$-$4$ bifurcations and then $1$-$q$, $q\ge5$, higher period ones~\footnote{This distinction is made through a known classification of ``strong'' $q<5$ and ``weak'' $q\ge5$ resonances in the study of area preserving maps~\cite{Arrowsmith1990}. In fact, the kicked $p$-spin model constrained to the vicinity of the poles is a family of generalized Hénon maps~\cite{MunozArias2021}, for which this distinction has been of utter importance~\cite{Dullin2000,Simo2009,Gonchenko2021}.}. Following previous studies~\cite{russomanno2017}, we will characterize the extension of the FTC phases in parameter space not only in the $\alpha(h)$ direction, but also in the direction of the interaction strength $\Lambda$. Notice that, for sufficiently large $\Lambda$, it is known that the kicked $p$-spin family transitions to chaos~\cite{MunozArias2021}, which will lead, unavoidably, to the system thermalizing in the long time limit. 

\subsection{Period doubling}
\label{subsec:period_doubling}

The $2T$-FTC phase emerging at the $1$-$2$ bifurcation in the vicinity of $\alpha_{\rm B} = \pi$ is the most prominent one for all even values of $p$, as the emergent symmetry agrees with the $\mathbb{Z}_2$ symmetry the Floquet system inherits from the time independent Hamiltonian. In particular, the case with $p=2$ recovers the FTC phase in the LMG first discussed in~\cite{russomanno2017}. 

In Fig. \ref{fig:peroid_doubling_tripling}a we show the results of the characterization of the $2T$-FTC phase for two example cases with $p=2,4$. Fig. \ref{fig:peroid_doubling_tripling}a.1.1 shows a snapshot of $f_Z(t)$, computed with the initial state $|\theta, \varphi\rangle = |\pi/5, 0\rangle$ in the long time limit. The period doubled oscillation is clearly seen for both the cases of $p=2$ and $p=4$, indicated by circles and triangles respectively. Fig. \ref{fig:peroid_doubling_tripling}a.1.2 shows the normalized power spectrum of $f_Z(t)$, displaying the clear peak at $\omega = \frac{2\pi}{2}$, accordingly to the subharmonic response of the system in the $2T$-FTC phase. The extension and robustness of these 2T-FTC phases are studied in the subsequent figures. Fig \ref{fig:peroid_doubling_tripling}a.2 shows the normalized mean adjacent spacing ratio $\tilde{r}(\alpha(h), \Lambda)$ in the vicinity of $\alpha_{\rm B} = \pi$ for systems with $p=2,4$, respectively. The nonwhite areas indicate the parameter region where the spectral statistics deviates from that of a Poisson distribution due to large degeneracies by pairs in the spectrum of $\hat{U}_F^2$. Finally, Fig. \ref{fig:peroid_doubling_tripling}a.3 shows the long-time average of the OTOC $\mathrm{F}^\infty_{Z,Z}(\Lambda, \alpha(h))$ for systems with $p=2,4$. The green area indicates $\mathrm{F}^\infty_{Z,Z} = 0$ and thus non-FTC phase, whereas nongreen area indicates $\mathrm{F}^\infty_{Z,Z} \ne 0$ thus FTC phase. 

At small values of $\Lambda$ the phase boundary of the FTC is essentially a straight line with a slope proportional to $\Lambda$, as can be seen in Fig. \ref{fig:peroid_doubling_tripling}a.2. At larger values of $\Lambda$, the shape of this boundary becomes more complicated, however it can be computed using the APM in the mean-field limit. We give an example for the system with $p=2$ in Appendix ~\ref{app:classical_equations}. The extent of the $2T$-FTC phase in the direction of $\Lambda$ is, in principle, unconstrained and there will always be a narrow strip of FTC phase around $\alpha_{\rm B}$ in the limit $\Lambda\to\infty$. This can be understood from the fact that the Floquet system at $\alpha(h) = \alpha_{\rm B}$ is fully integrable and never transitions to chaos. In fact, the largest Lyapunov exponent in the limit of strong chaos (thermal regime), is given by $\lambda_{+}(\alpha(h),\Lambda,p) = \ln\left[\Lambda(p-1)\sin(\alpha(h)) \right] - (p-1)$ (see~\cite{MunozArias2021} for details). Hence, as $\alpha(h)\to\pi$ one requires $\Lambda \to \infty$ in order to give $\lambda_+ > 0$. This fact is illustrated in Fig. \ref{fig:peroid_doubling_tripling}a.2,a.3 (also Fig. \ref{fig:classical_phase_diagram}) where the region of the FTC phase appears to extent indefinitely in the $\Lambda$ direction. This is a feature only of the $2T$-FTC phase and its emergence in the vicinity of $\alpha_{\rm B} = \pi$, and it is not present in the other FTCs present in this system.

\subsection{1-3 bifurcation}
\label{subsec:predio_tripling}
As discussed in Sec. \ref{sec:reso_reso_cond_bifus}, we can go beyond period doubling FTCs by setting $\alpha_B=\frac{2\pi}{3}$ leading to a $1$:$3$ resonance and identifying for which models a significant period-tripling bifurcations takes place, i.e. for which model $\frac{2\pi}{3}\in\mathcal{B}_{\rm bifus}^{(p)}$. The first of those models corresponds to $p=3$, the second and third ones have $p=5,6$, respectively. We illustrate the characteristic period tripling behavior of this phase using the systems with $p=3,6$ in Fig. \ref{fig:peroid_doubling_tripling}b. In Fig. \ref{fig:peroid_doubling_tripling}b.1.1 we show a snapshot of $f_Z(t)$ computed with the initial state $|\phi_0\rangle = |0,0\rangle$, where dots and triangles show $p=3,6$ respectively. The period tripling behavior is manifested in the normalized power spectrum of $f_Z(t)$ in Fig. \ref{fig:peroid_doubling_tripling}b.1.2, where the peak is locked at $\omega = \frac{2\pi}{3}$ for both cases.

The extension of the phase in parameter space is investigated using the normalized average level spacing ratio $\tilde{r}(\Lambda, \alpha(h))$ and the long time average of the OTOC $\mathrm{F}^\infty_{Z,Z}(\Lambda, \alpha(h))$. In Fig. \ref{fig:peroid_doubling_tripling}b.2 we show $\tilde{r}(\Lambda, \alpha(h))$ in the vicinity of the bifurcation point, where white areas indicates Poisson statistics and nonwhite areas indicates deviations from Poisson statistics, which we connect to presence of FTC behavior. Notice that the FTC phase for the system with $p = 3$ in Fig. \ref{fig:peroid_doubling_tripling}b.2.1 is highly constrained, as chaos emerges around this value of $\alpha(h)$ fairly rapidly ~\cite{MunozArias2021}, erasing any footprint of the locked subharmonic periodic behavior. On the other hand, for the system with $p=6$ the phase is wider in both, the $\alpha$ and $\Lambda$ directions. For this case, the system gains stability thanks to the $\mathbb{Z}_2$ symmetry of the even $p$ spin systems, and the period-3 bifurcation is double, thus being structurally similar to a $1$-$6$ bifurcation. The FTC phase extents to larger values of $\Lambda$ since higher values of $p$ have a delayed emergence of global chaos~\cite{MunozArias2021}. The long time average of the OTOC in Fig.~\ref{fig:peroid_doubling_tripling}b.3 shows very good agreement with the results of $\tilde{r}$ for the $3T$-FTC phase. Here, green indicates nonFTC behavior and nongreen indicates FTC behavior. 

\subsection{1-4 bifurcation}
\label{subsec:period_quadripling}
\begin{figure*}[!t]
 \centering{\includegraphics[width=0.98\textwidth]{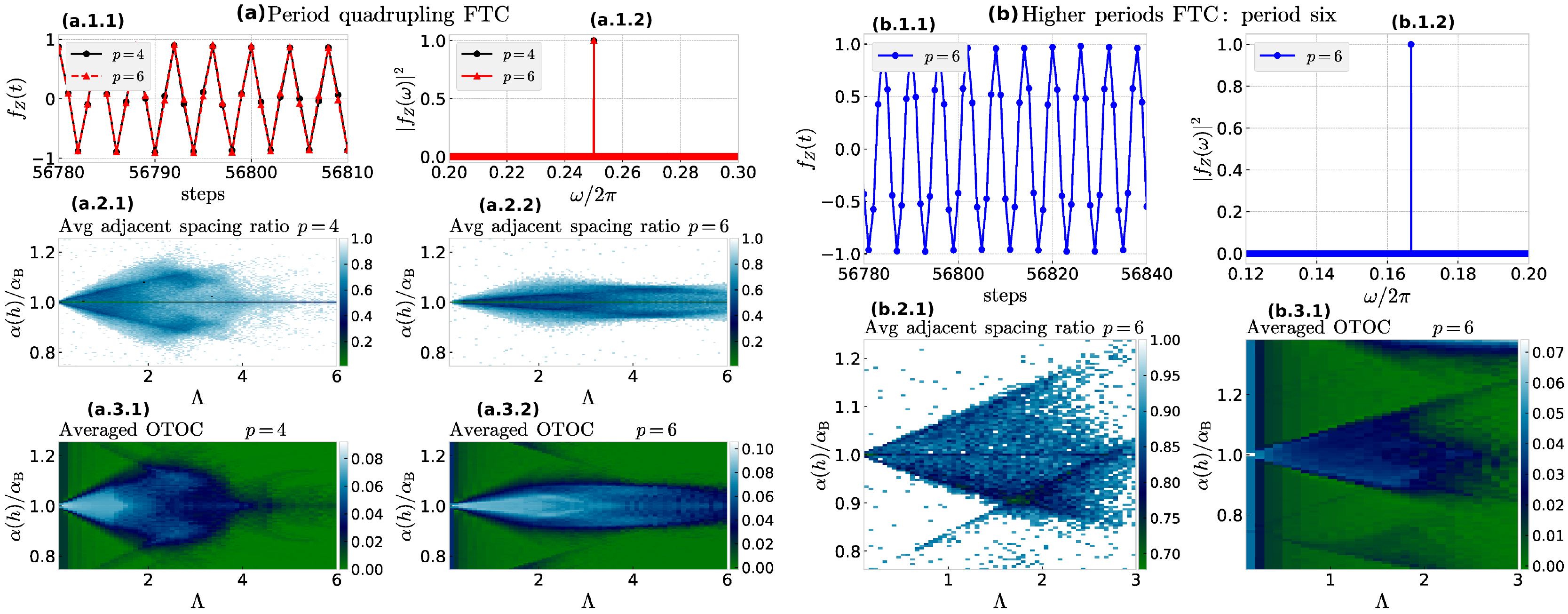}}
\caption{Characterization of the $4T$-FTC \textbf{(a)}, and $6T$-FTC \textbf{(b)} phases in our family of models. \textbf{(a.1) - (b.1)} Snapshot of $f_Z(t)$ in the long time and its normalized power spectrum. A clear subharmonic response with $\omega = \frac{1}{4}, \frac{1}{6}$ is observed in \textbf{(a)} and \textbf{(b)}, respectively. \textbf{(a.2) - (b.2)} $\tilde{r}$ as a function of $\alpha(h)$ and $\Lambda$ in the vicinity of $\alpha_{\rm B} = \frac{2\pi}{4}$ and $\alpha_{\rm B} = \frac{2\pi}{6}$, respectively. White indicates Poisson statistics, nonwhite indicates some degree of quartets/sextets clustering of levels and indicates the region of parameters where the system behaves as a $4T$-FTC and $6T$-FTC, respectively. \textbf{(a.3)-(b-3)} $\mathrm{F}^\infty_{Z,Z}$ as a function of $\alpha(h)$ and $\Lambda$ in the vicinity of $\alpha_{\rm B} = \frac{2\pi}{4}$ and $\alpha_{\rm B} = \frac{2\pi}{6}$, respectively. Green indicates $\mathrm{F}^\infty_{Z,Z} = 0$, nongreen indicates $\mathrm{F}^\infty_{Z,Z} \ne 0$, hence the region of parameters where the system behaves as a $4T$-FTC and a $6T$-FTC, respectively. Parameters are: $N = 1024$ in \textbf{(a.2)} and \textbf{(b.2)}, and $N = 256$, $T_{\rm max} = 16000$ in \textbf{(a.3)} and \textbf{(b.3)}. $h = 0.05$, $\Lambda = 0.7$, $T_{\rm max} = 60000$, $N = 1024$ in \textbf{(a.1)} and $h = 0.05$, $\Lambda = 1.0$, $T_{\rm max} = 60000$, $N = 1024$ in \textbf{(b.1)}.}
\label{fig:period_quadrupling_higher}
\end{figure*}
Following the discussion in Sec. \ref{sec:reso_reso_cond_bifus}, models where $\frac{2\pi}{4}\in\mathcal{B}_{\rm bifus}$ host a $4T$-FTC phase in the vicinity of $\alpha_{\rm B} = \frac{2\pi}{4}$. We fix the angle of the drive at this value and investigate the emergence of this phase with two exemplary systems with $p=4,6$. Fig \ref{fig:period_quadrupling_higher}a.1.1 shows, for the initial state $|\theta, \varphi\rangle = |\pi/5, 0\rangle$, a snapshot of $f_z(t)$ in the long time limit, revealing the period quadrupling behavior (circles and triangles indicate $p = 4,6$, respectively). Correspondingly, Fig. \ref{fig:period_quadrupling_higher}a.1.2 shows the normalized power spectrum of $f_Z(t)$, displaying a clear peak at the subharmonic frequency $\omega = \frac{2\pi}{4}$ in both cases.

The extension of this FTC phase in parameter space is investigated with both the normalized average level spacing ratio $\tilde{r}(\Lambda, \alpha(h))$ and the long-time average of the OTOC $\mathrm{F}^\infty_{Z,Z}(\Lambda, \alpha(h))$. The normalized spectral statistics indicator shows a region in parameter space associated with the $4T$-FTC phase, which grows in the $\Lambda$ direction as we increase the value of $p$, compare Fig. \ref{fig:period_quadrupling_higher}a.2.1 ($p=4$) and Fig. \ref{fig:period_quadrupling_higher}a.2.2 ($p=6$), here white indicates Poisson statistics and nonwhite indicate a value deviating from Poisson statistics which we associate with some degree of level clustering by quartets. The long time average of the OTOC is shown in Fig. \ref{fig:period_quadrupling_higher}a.3 for the systems with $p=4,6$, respectively. Here green indicates $\mathrm{F}^\infty_{Z,Z} = 0$ and nongreen indicates $\mathrm{F}^\infty_{Z,Z}\ne0$. Notice that $\mathrm{F}^\infty_{Z,Z}$ agrees with $\tilde{r}$ on the region they associate with signatures of a $4T$-FTC. We point out that these two indicators identify very well the finite size phase diagram of our FTC phases, regardless of the period of the phase.

An important issue arises here. Could we have identified a $4T-FTC$ for a system with $p=2$? As we mentioned in the Sec. \ref{sec:intro} and Sec. \ref{sec:reso_reso_cond_bifus}, area preserving maps naturally develop resonances, a simple way of identifying them is via the resonance conditions associated with a delta-kicked Hamiltonian. Thus, one could define FTC phases (or, rather, their precursors) using those resonances. However, the requirement of a global $\mathbb{Z}_q$ dynamical symmetry is usually not satisfied. This is the case for $p=2$ when one tries to define a $4T$-FTC at at the resonance condition $\alpha_{\rm B} =  \frac{2\pi}{4}$. We discuss this case in more detail in Appendix \ref{app:absence_FTC}. 

\subsection{Higher-order bifurcations}
\label{subsec:higher_period}
So far we have discussed FTC phases up to period quadrupling. However, higher period FTC phases are accessible provided we work with a system with large enough $p$-body interactions. A given kicked $p$-spin will always have a $1$-$p$ bifurcation at $\alpha_{\rm B} = \frac{2\pi}{p}$, whose vicinity can be used to define a $pT$-FTC phase. Additional $qT$-FTC phases with $q<p$ could be defined, in the vicinity of the appropriate value of $\alpha_{\rm B}$, if $\alpha_{\rm B} = \frac{2\pi}{q} \in \mathcal{B}_{\rm bifu}$.

Let us consider the system with $p=6$, whose lower period FTC phases have been analyzed in the previous subsections. We focus on values of $\alpha(h)$ in the vicinity of $\alpha_{\rm B} = 2\pi/6$, point at which the corresponding mean-field system has a $1$-$6$ bifurcation. The response of the system is characterized using $f_Z(t)$ for an initial state of the form $|\theta, \varphi\rangle = |\pi/10, 0\rangle$. Fig.~\ref{fig:period_quadrupling_higher}b.1.1 shows a snapshot in the long time limit of this response, where a clear $6$-periodic oscillation can be observed. To complement this observation, we look at the normalized power spectrum of the $f_Z(t)$ signal in Fig.~\ref{fig:period_quadrupling_higher}b.1.2, showing a clear peak at the frequency $\omega = 2\pi/6$, signaling a $6T$-FTC. 

To investigate the extension of this FTC phase in parameter space we look at both, the average adjacent level spacing ratio $\tilde{r}(\Lambda, \alpha(h))$ and the long-time average of the OTOC $\mathrm{F}^\infty_{Z,Z}(\Lambda, \alpha(h)$. In Fig.~\ref{fig:period_quadrupling_higher}b.2.1 the spectral statistics indicator is shown, where nonwhite areas indicate deviation from Poisson statistics providing evidence for the level clustering by sextets, a signature of the onset of the $6T$-FTC phase. In Fig.~\ref{fig:period_quadrupling_higher}b.3.1 the long time average of the OTOC is shown. Here, green indicates $\mathrm{F}^\infty_{Z,Z} = 0$ and nongreen indicates $\mathrm{F}^\infty_{Z,Z} \ne 0$ signaling a nonFTC and FTC phases, respectively. 

We close the analysis of finite-size systems by stressing that the two novel indicators proposed here, one based on the spectral statistics of the appropriate power of the Floquet operator and the other based on the behavior of the OTOC, have shown to provide excellent characterizations of the nonFTC-FTC phases. These two indicators are also complementary, as the first one is purely kinematic and the second one purely dynamic, providing then interesting tools for the characterization of FTC phases in systems beyond the $p$-spin models studied here.

\subsection{Emergent FTC phases in the thermodynamic limit}
\begin{figure}[!t]
 \centering{\includegraphics[width=0.48\textwidth]{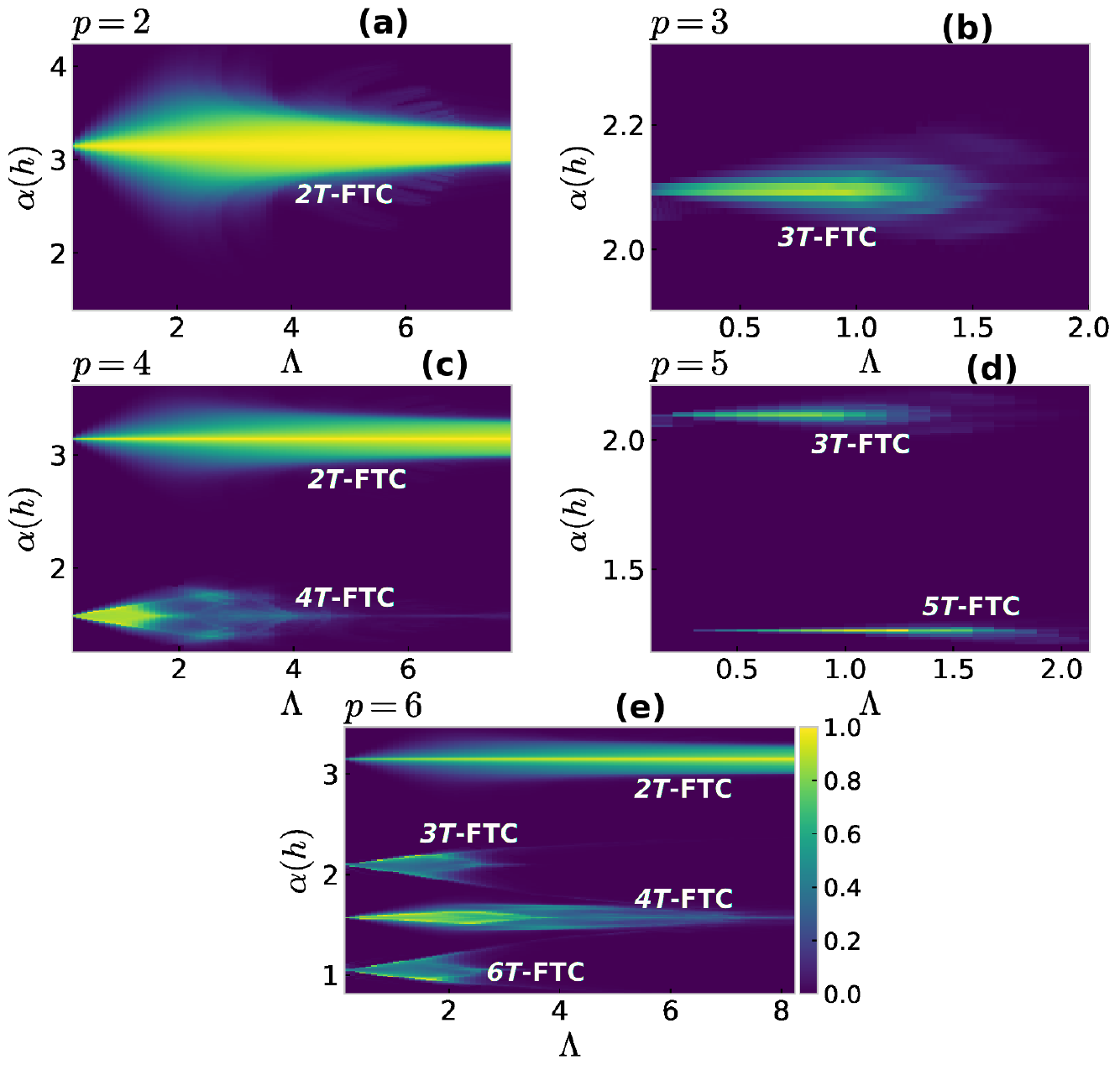}}
\caption{Phase diagrams of the different FTC phases in the mean-field limit constructed using $\mathcal{G}(\Lambda,\alpha(h))$ defined in Eq. (\ref{eqn:class_phase_diagram}), for the kicked $p$-spin systems with $p=2,3,4,5,6$ in panels \textbf{(a-e)}, respectively. Notice the correspondence between the observed phases and the set of resonant-bifurcation points $\mathcal{B}_{\rm bifu}^{(p)}$.}
\label{fig:classical_phase_diagram}
\end{figure}
The results in previous subsections characterize the emergent FTC phases for finite system sizes. In this subsection we will construct the phase diagram of the different FTC phases directly in the thermodynamic limit, for kicked $p$-spin models up to $p=6$. For these models, this can be done simply by solving numerically the nonlinear classical equations of motion for the magnetizations $(X,Y,Z)$. From this, we can compute the measure $\mathcal{G}(\Lambda, \alpha(h))$ introduced in Sec.~\ref{subsec:measure_for_classical_phase_diagram}. In the following, we show results where the phase space average was done using an uniform grid on the sphere with $14000$ points, and evolution times going up to $T_{\rm max} = 16000$.

Results are shown in Fig. \ref{fig:classical_phase_diagram}. In all cases with even $p$ the $2T$-FTC phase emerging around $\alpha_{\rm B} = \pi$ is robust in the $\Lambda$ direction. As pointed out previously, for this angle of the drive the system never transitions to chaos, and thus the system avoids thermalization in the long-time limit (see discussion in Sec. \ref{subsec:period_doubling}). The $4T$-FTC phase in the system with $p=4$ exists up to $\Lambda\sim5$ as it is then that the system becomes globally chaotic~\cite{MunozArias2021}. On the other hand, as we increase $p$ the $4T$-FTC phase gains in extension along the $\Lambda$ direction, since higher values of $p$ delay the emergence of global chaos in this family of models~\cite{MunozArias2021}.

Some other important observations can be drawn from these results. For the models with odd values of $p$, FTC phases tend to be highly constrained (notice the extent of the horizontal axis in Figs. \ref{fig:classical_phase_diagram} (b) and (d) with respect to the rest). For instance, for $p=3$ the model hosts a $3T$-FTC phase around $\alpha_{\rm B} = \frac{2\pi}{3}$, however the poles at the bifurcation point are unstable and chaos emerges fairly rapidly for this model around this value of the angle of the drive. In fact as seen in Fig. \ref{fig:classical_phase_diagram}b the phase only goes up to $\Lambda~\sim1.5$. The $3T$-FTC phase in the model with $p=5$ is essentially identical to that in the model with $p=3$, whereas the $5T$-FTC is narrower in the direction of $\alpha$. This phenomenon keeps occurring as $p$ becomes larger in the models with odd $p$'s. Finally, the $3T$-FTC phase in the system with $p=6$ is structurally similar to the $6T$-FTC phase, see Fig. \ref{fig:classical_phase_diagram}e. This is due to the fact that the APM of the kicked $p$-spin, $\mathcal{A}[\mathbf{X}_l;\alpha(h), \Lambda, p]$, is a double reversible map for models with even $p$~\cite{Mackay1993,MunozArias2021}, in other words, the inherited $\mathbb{Z}_2$ symmetry imposes special constraints to bifurcations which have odd period, in this case they are \textit{double}, meaning that one observes the emergence of two $1$:$3$ resonances looking structurally identical to a single $1$:$6$ resonance. Finally, we highlight the resemblance between the finite size phase diagrams constructed with the spectral statistics indicator and the long-time average of the OTOC, and shown in Fig.~\ref{fig:peroid_doubling_tripling} and Fig.~\ref{fig:period_quadrupling_higher}, with those constructed for the systems in the, $N\to\infty$, thermodynamic limit, discussed in the present subsection and shown in Fig. \ref{fig:classical_phase_diagram}.

\section{Time Crystal Switching}
\label{sec:time_crystal_switching}
\begin{figure}[!t]
 \centering{\includegraphics[width=0.5\textwidth]{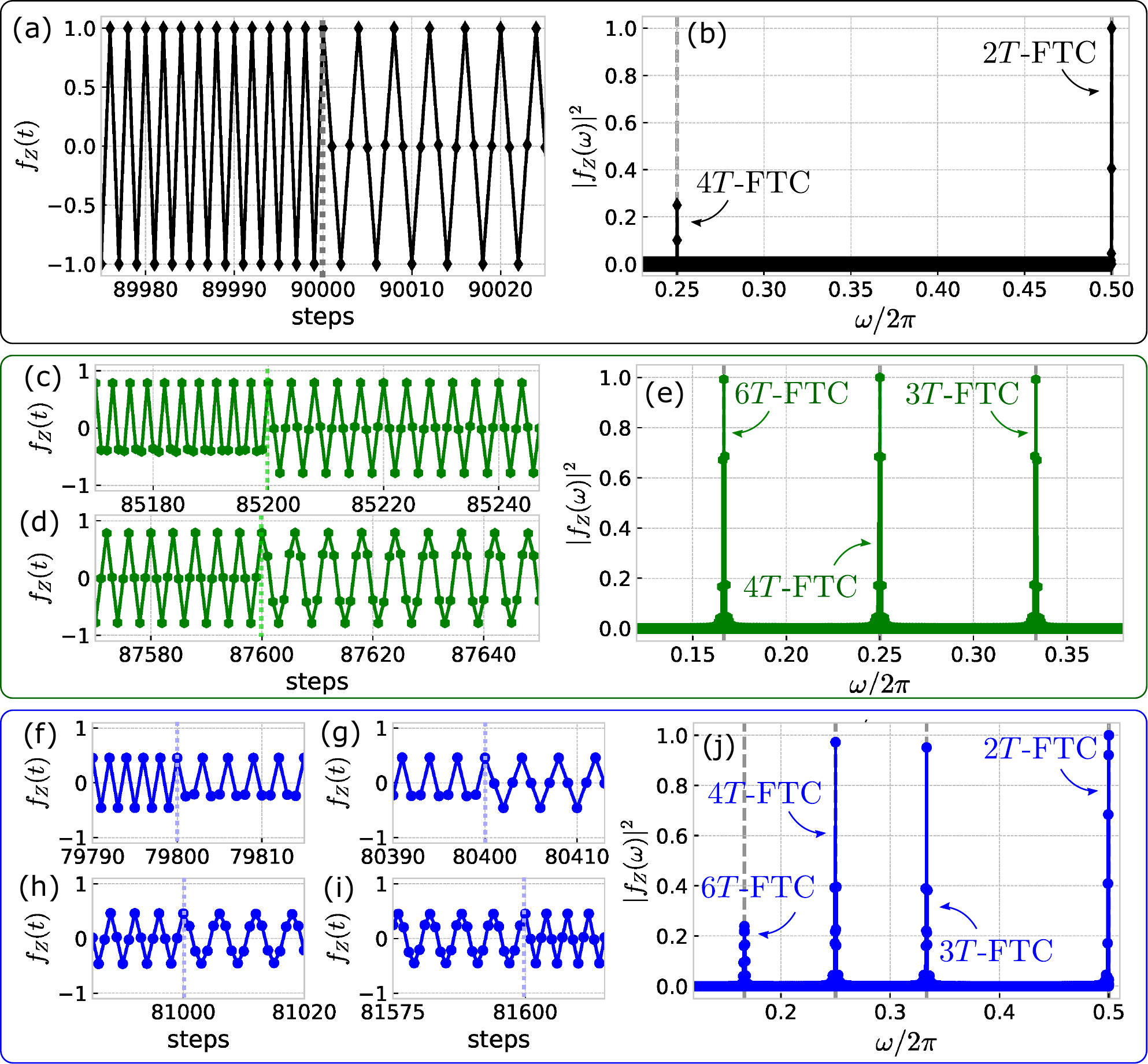}}
\caption{\textbf{(a,b)} Time crystal switching in the system with $p=4$. \textbf{(a)} Example of $f_Z(t)$ at the switching moment between a $2T$-FTC and a $4T$-FTC. \textbf{(b)} Normalized power spectrum of $f_Z(t)$. We can immediately identify the two subharmonic frequencies which take part in the system response, $\omega = \frac{1}{2}$ and $\omega = \frac{1}{4}$. \textbf{(c-e)} Example of time crystal switching in the system with $p=6$, involving a $3T$-FTC, a $4T$-FTC, and a $6T$-FTC. \textbf{(c,d)} Examples of $f_Z(t)$ showing the switching between $3T$-FTC to $4T$-FTC and between $4T$-FTC and $6T$-FTC. \textbf{(d)} Normalized power spectrum of $f_Z(t)$ displaying three clear peaks at the subharmonic frequencies which take part in the system response. \textbf{(f-i)} Examples of switching in the system with $p=6$, showing switching between $2T$-FTC and $3T$-FTC, between $3T$-FTC and $4T$-FTC, between $4T$-FTC and $6T$-FTC, between $6T$-FTC and $4T$-FTC, respectivley. \textbf{(j)} Normalized power spectrum of $f_Z(t)$ showing four clear peaks at the sunharmonic frequencies appearing in the system response. The dashed vertical lines are guides for the eye and they signal the different subharmonic frequencies of the FTC phases we switch between. Paramters are: $h = 0.02$ in \textbf{(a,b)} and $h = 0.01$ in \textbf{(e-j)}, $\Lambda = 0.7$, and $N = 1024$.}\label{fig:tc_switching}
\end{figure}

In this section we propose a family of protocols that enable to switch the system response between different FTC phases by quenching a parameter of the Hamiltonian in time, but in a non-periodic way. In previous sections we considered the rotation angle $\alpha_B$ in Eq. (\ref{eqn:floquet_ope}) as the parameter that tunes between different FTC phases for a given value of $p$. In particular, we discussed how setting $\alpha_B$ at or near one of the values in the set $\mathcal{B}_{\rm bifu}^{(p)}$ lead to a particular FTC phase. As discussed in the previous sections, the set $\mathcal{B}_{\rm bifu}^{(p)}$ has more than one element for all $p>3$. Physically, this implies that for a given $p$ the same family of initial states (localized within the ferromagnetic well), one obtains a robust subharmonic dynamical response of the system with a period that depends only on the choice of $\alpha_B$. Interestingly, from a physical point view, $\alpha_B$ can be regarded as the strength of an external field, which is applied to the system as a train of periodic pulses. So, we propose a control scheme in which this parameter changes in time in order to switch, \textit{on the fly}, the subharmonic response of the system from one phase to the other. The easiest protocol that can achieve this is a sudden change, or ``quench''  from, say $\alpha_B^{(1)}$ to  of $\alpha_B^{(2)}$ happening at a some time $t=mT$, with arbitrary $m\in\mathbb{N}$ and $\alpha_B^{(i)}\in\mathcal{B}_{\rm bifu}^{(p)}$.  We call this scheme Time Crystal Switching. 

For concreteness, consider $p=4$ which is the simplest case that can host this behavior. As discussed in previous sections and illustrated in Fig. \ref{fig:reso_spheres}, in this case we have that $\mathcal{B}_{\rm bifu}^{(p=4)}=\{\pi,\frac{\pi}{2}\}$, leading to a period-doubling and period-quadrupling FTC phase respectively. The Time Crystal Switching protocol involves preparing an initial fully polarized state with extensive support inside the region enclosed by the separatrix, driving the system with a train of $m$ pulses at an angle $\alpha^{(1)}_{\rm B}=\pi$, and then quenching the field to produce an angle $\alpha^{(2)}=\frac{\pi}{2}$. The resulting dynamics is shown in Fig. \ref{fig:tc_switching} (a) and (b) where the sudden change of the subharmonic response of the system from $\omega = \frac{2\pi}{2}$ to $\omega = \frac{2\pi}{4}$ can be clearly seen. 

As we increase the value of $p$, the richness of the available protocols increases substantially. We illustrate this for the case of $p=6$, in which the set of available values of $\alpha_B$ hosting FTC phases is $ \mathcal{B}_{\rm bifu}^{(p=6)}=\left\{\pi,\frac{2\pi}{3},\frac{\pi}{2},\frac{\pi}{3}\right\}$. We can then consider a variety of possible protocols. In a first example, we take the usual initial stretched state along the $z-$axis $\Ket{\psi_0}=\Ket{S,S}$ and choose to quench $\alpha_B$ in time in the sequence $\frac{2\pi}{3}\rightarrow\frac{2\pi}{4}\rightarrow \frac{2\pi}{6}$. The results can be seen in Fig. \ref{fig:tc_switching}c-e, where we observe how the system switches from a $3T$-FTC to a $4T$-FTC and finally to a $6T$-FTC. As a final illustration, we consider a more intricate protocol where, for the same initial state, we now quench $\alpha_B$ in the sequence:  $\pi \rightarrow \frac{2\pi}{3} \rightarrow \frac{2\pi}{4} \rightarrow \frac{2\pi}{6}$. Again, we observe a clear transition between different FTCs, first with period 2 to 3, then 3 to 4, and finally 4 to 6. Note that, although the switching process introduces a small amount of modulation and thus some heating, reflected in the emergence of small broadening of each of the subharmonic peaks in Fig. \ref{fig:tc_switching}b,e,j, the system remains locked to the appropriate frequency of each of the FTC phases it is switching into.

We stress that, even though the feasibility of this protocol follows naturally from the description of FTC phases in terms of classical area-preserving maps, from a physical point of view it is enabled by a set of unique properties of the kicked $p$-spin models, which we have described extensively in this work: (i) increasing the degree of the interaction $p$ leads to many different FTC phases with different periods $mT$, with $m\leq p$, (ii) the subharmonic response is seen for an extensive family of initial states, which is roughly independent of the particular FTC phase, and determined by the equilibrium properties of the underlying $p-$spin Hamiltonian, and (iii) the parameter controlling the period of the FTC phase, $\alpha_B$, can be regarded as produced by an external magnetic field,  which in principle can be tuned in time in a way that is independent of the details of the interacting many-body system. We consider that these time crystal switching protocols are just a particular example of a more general class of control schemes in Floquet systems which could lead to novel responses beyond the usual FTC phases.


\section{Conclusions and outlook}
\label{sec:final_remarks}
In this work we have shown that periodically driven $p$-spin models can host a wide variety of mean-field Floquet Time Crystal (FTC) phases showing robust subharmonic responses with period $mT$, where $m$ is as low as 2 and as high as $p$, and T is the period of the drive.
The dynamics of these models can be treated exactly in the thermodynamic limit via its mean-field description, which takes the form of an area-preserving map (APM), i.e. the classical dynamical systems analog of a Floquet map. We have identified the precursor of the subharmonic response of the system as the existence of a resonance in the APM, that is, a parameter regime in which an integer number of periods of the drive fits exactly in one natural cycle of the undriven system. Then, we have discussed how bifurcations at these resonance points explain the major structural changes in phase space which accompany the emergence of new dynamical symmetries leading to an FTC phase. In particular, the emergence of a global $q$-fold rotational symmetry of phase space can be considered as one of the defining characteristics of FTC phases~\cite{Else2020}. Finally, we have shown that the degree of the multi-body interaction $p$ determines which of these resonant bifurcations will ultimately lead to a proper FTC phase in the driven system, highlighting the key role played by of the multi-body interaction in the studied mean-field FTC phases.


Using the insight provided by the mean-field analysis, we have predicted and extensively described FTC phases in driven $p$-spin models in the quantum regime, and showed that the subharmonic response is robust to parameter variations even for finite system sizes. We have shown that systems with $p>3$ can host several FTC phases, and that phases with higher periods are less robust to parameter variations. In all cases, however, we have identified a finite parameter regime for which the phases survive in the thermodynamic limit. 

In the quantum regime, we have proposed a number of static and dynamical indicators to characterize the emergence of FTC phases. On one hand, we have analyzed the emergence of clustering in the spectrum of Floquet eigenphases as signature of the emerging eigenstate ordering. Furthermore, we have argued that in the mean-field models this ordering can be identified with a clustering ESQPT of the effective Hamiltonian associated with $\hat{U}_F^q$. Conversely, we have shown that the nonFTC-FTC transition can be described as a dynamical quantum phase transition (DQPT). Following recent works ~\cite{Heyl2018,BurkaDag2019,Wang2019}, we used the long-time average of out of time order correlators to detect such transition. 

The results presented here open several several different avenues of future research. One of them is to explore in more detail the connection between multi-body interactions and higher-order FTC phases, beyond the mean-field regime. A path forward in this direction is to consider many-body models with finite-range multi-body interactions. Even without driving, breaking the natural permutation symmetry present in the $p$-spin models will break the integrability of the system, and most studies so far have been focused on $p=2$, which has a natural connection to ion-trap quantum simulators \cite{Islam2011,Britton2012,Zhang2017,Zunkovic2018}. For the case with driving, not much is known even for $p=2$. Going beyond the mean-field regime would also enable to explore the relation between the behavior observed here and so-called many-body resonances reported in Ref.~\cite{bukov2016}, which correspond to parameter regimes in which an otherwise ergodic driven quantum system fails to thermalize and shows long-lived temporal correlations. \\

As a side note, even though in this paper we have been concerned with breaking of the discrete time-translation invariance, the relation we have found between the large-period FTCs and the order of the many-body interaction bears an interesting resemblance to previous results on \textit{continuous} Time Crystals. In Ref.~\cite{kozin2019}, the authors found that some long-range interacting spin-1/2 models could lead to the desired symmetry breaking (bypassing the no-go theorem previously proven in~\cite{watanabe2015}). The required Hamiltonian is intrinsically non-local and contains $p$-body interactions, however $p\sim N$ in that case, as opposed to $p\ll N$ in the cases discussed in the present work. 

Finally, another natural extension of the present work is to explore more general control protocols in Floquet systems which can host more than one FTC phase as a control parameter is varied. In this work we have explored a simple protocol where the control parameter is quenched between the values corresponding to different phases. However, more general schemes could be deviced; for instance a slow, quasi-adiabatic passage between those two values. These protocols could lead to interesting responses of the system beyond the time crystal switching shown in Sect. \ref{sec:time_crystal_switching}. 

\acknowledgments
The authors are grateful to Ivan H. Deutsch for insightful discussions during the development of this work. The authors also gratefully acknowledge Norman Yao for insightful discussions and clarifications on the general idea of time crystal phases in out-of-equilibrium quantum matter.  We also extend our acknowledgments to Ricardo Puebla Antunes for insightful discussions on excited state quantum phase transitions and dynamical quantum phase transitions, to  Ceren B. Da\ifmmode \breve{g}\else \u{g}\fi{} for her insights in the use of OTOCs for the dynamical detection of quantum phases and quantum phase transitions, and to Austin K. Daniel for his insights into discrete symmetry groups. This work was supported by NSF Grants No. PHY-2011582, No. PHY-1820758 and Quantum Leap Challenge Institutes program, Award No. 2016244. This material is based upon work supported by the U.S. Department of Energy, Office of Science, National Quantum Information Science Research Centers, Quantum Systems Accelerator (QSA).

\appendix
\section{Classical Equations of Motion}
\label{app:classical_equations}
In this appendix we give explicit expressions for the equations of motion of the classical flow associated with the undriven system and the area preserving map associated with the Floquet system.

The phase space flow associated with Eq. (\ref{eqn:p_spin_hamil}) can be obtain from the Heisenberg equations of motion of $\hat{\mathbf{S}}$, and in the limit $S \to \infty$ assuming all correlations factor, that is, $\langle\hat{A}\hat{B} \rangle = \langle\hat{A}\rangle \langle \hat{B}\rangle$. After those steps, the equations of the flow are 
\begin{subequations}
\label{eqn:flow_equations}
\begin{align}
\frac{d X}{dt} &= \Lambda Z^{p-1}Y, \\
\frac{d Y}{dt} &= hZ - \Lambda Z^{p-1}X, \\
\frac{d Z}{dt} &= -h Y.
\end{align}
\end{subequations}
The classical equations associated with the kicked evolution in Eq. (\ref{eqn:kicked_p_spin_uni}) are obtained from the classical limit of the Heisenberg evolution of $\hat{\mathbf{S}}$, that is, $\hat{\mathbf{S}}_{l+1} = \hat{U}_F^{\dagger l}\hat{\mathbf{S}}_l \hat{U}_F^{l}$, and they are given by 
\begin{subequations}
\label{eqn:classical_map}
\begin{align}
X_{l+1} &= \cos\left(\Lambda Z^{p-1}_l \right)X_l - \sin\left(\Lambda Z_l^{p-1} \right) Y_l, \\
Y_{l+1} &= \left[ \sin\left( \Lambda Z_l^{p-1} \right)X_l + \cos\left(\Lambda Z_l^{p-1} \right)Y_l \right]\cos\left(\alpha(h)\right) \nonumber \\ 
&- \sin\left(\alpha(h)\right)Z_l, \\ 
Z_{l+1} &= \left[ \sin\left( \Lambda Z_l^{p-1} \right)X_l + \cos\left(\Lambda Z_l^{p-1} \right)Y_l \right]\sin\left(\alpha(h)\right) \nonumber \\
&+ \cos\left(\alpha(h)\right)Z_l.
\end{align}
\end{subequations}
In order to find the bifurcation points of the poles $X \pm 1$, we evaluate the tangent map of Eq. (\ref{eqn:classical_map}) at the fixed point. We find two different situations, when $p=2$ the eigenvalues read
\begin{multline}
\label{eqn:eig_vals_2p}
\mathcal{A}_{\pm}(\Lambda,\alpha(h)) = \frac{\pm \Lambda \sin(\alpha(h))+2\cos(\alpha(h))}{2} \\ \pm \frac{1}{2}\sqrt{(\mp \Lambda\sin(\alpha(h)) - 2\cos(\alpha(h)))^2 - 4},
\end{multline}
for all the other models with $p>2$, the eigenvalues read 
\begin{equation}
\label{eqn:eig_vals_tother_p}
\mathcal{A}_{\pm}(\alpha(h)) = e^{\pm i \alpha(h)},
\end{equation}
thus the models with $p>2$ the poles undergo a $d$-$q$ bifurcation anytime $\alpha(h) = \frac{2\pi d}{q}$, with $d$, $q$ relative primes and $q\ge2$. Further details of stability calculations can be found in~\cite{MunozArias2021}.

\begin{figure}[!t]
 \centering{\includegraphics[width=0.47\textwidth]{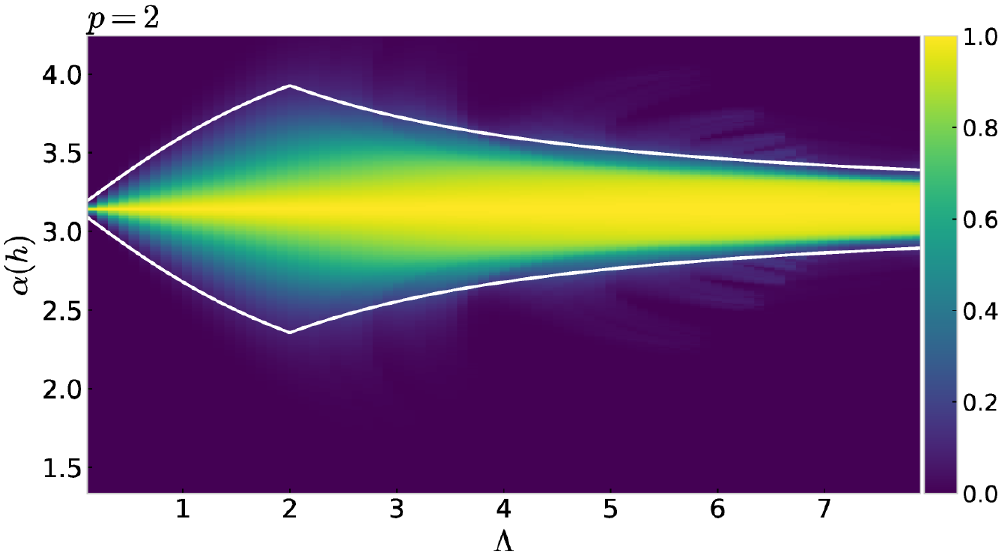}}
\caption{Classical phase diagram for the $2T$-FTC in the $p=2$ system. The white continuous line shows the phase boundary constructed using Eq.~(\ref{eqn:phase_boundary_2p}). This expression provides an excellent approximation to the phase boundary.}\label{fig:phase_diagram_boundary}
\end{figure}

Before finishing this appendix let us comment on an additional calculation which the eigenvalues above, and the resonance Hamiltonian in Eq. (\ref{eqn:resonance_hamiltonian}), allow us to do. We can estimate the boundary of the FTC phases in the thermodynamic limit. Let us illustrate this with the case of the $2T$-FTC in the $p = 2$ system. For this particular FTC phase we take $\alpha(h) = \pi + h$ and look for conditions such that in Eq. (\ref{eqn:eig_vals_2p}) is a real number, that is, the fixed point is hyperbolic, hence its separatrices enclosed a $1$:$2$ resonance. We find 
\begin{equation}
h_{2T {\rm -FTC}}^{(p=2)} = {\rm arctan}\left(\frac{4\Lambda}{4 - \Lambda^2}\right).
\end{equation}
We can now write the boundary of this phase as given by the curves 
\begin{equation}
\label{eqn:phase_boundary_2p}
\alpha(\Lambda) = \pi \pm \frac{1}{2}{\rm arctan}\left(\frac{4\Lambda}{4 - \Lambda^2}\right),
\end{equation}
where we have included a factor of $\frac{1}{2}$ accounting for the fact that time is being measured in steps of two. To illustrate how well the above expressions approximates the boundary of the $2T$-FTC phase  in the $p=2$ system, we show in Fig. \ref{fig:phase_diagram_boundary} the corresponding phase diagram with Eq.~(\ref{eqn:phase_boundary_2p}) on top (withe solid line). 

\section{Spinodal, ground state critical and dynamical critical points in the $p$-spin family}
\label{app:spinoal_and_criti} 

The starting point for the computation of spinodal and critical points is the semiclassical energy function in Eq. (\ref{eqn:semiclassical_energy}) constrained to the $x-z$ plane, as it is in this plane where it develops new extreme values. We thus have 
\begin{equation}
\label{eqn:constrained_semi_energy}
E(Z;h,\Lambda,p) = -h\sqrt{1-Z^2} - \frac{\Lambda}{p}Z^p.
\end{equation}
This equation has two extreme points at $X = \pm1$. New extreme points are solutions of the implicit equation 
\begin{equation}
\label{eqn:implciit_equation}
Z = W(p) Z^{p-1}\sqrt{1 - Z^2},
\end{equation}
where $W(p) = \frac{\Lambda}{h}$. For the calculation of each of the three points of interest, we need to find a complementary equation to Eq. (\ref{eqn:implciit_equation}), such that the resulting system of coupled equations has a solution. 

In the case of the spinodal point we define 
\begin{equation}
\label{eqn:aux_spinodal_1}
g(Z;W,p) = W(p) Z^{p-1}\sqrt{1 - Z^2},
\end{equation}
and notice that the spinodal point corresponds with a bifurcation point of the above equation, thus the complementary equation is given by $\frac{dg}{dZ} = 1$, that is
\begin{equation}
\label{eqn:aux_spinodal_2}
(p-1)W(p) Z^{p-2}\sqrt{1-Z^2} - \frac{W(p) Z^p}{\sqrt{1 - Z^2}} = 1.
\end{equation}
Additionally, from Eq. (\ref{eqn:implciit_equation}) one has 
\begin{equation}
\label{eqn:u_val}
W(p) = \frac{1}{Z^{p-2}\sqrt{1 - Z^2}},
\end{equation}
plugging this value in Eq. (\ref{eqn:aux_spinodal_2}) and solving for $Z$, then using the found value of $Z$ to solve for $W$, we find the solutions 
\begin{equation}
\label{eqn:spino_solutions}
Z_{\rm spino} = \sqrt{\frac{p-2}{p-1}}, \enspace W_{\rm spino}(p) = \sqrt{\frac{(p-1)^{p-1}}{(p-2)^{p-2}}}.
\end{equation}
which is the same vale as the one in Eq. \ref{eqn:spino_point} in the main text.

For the ground state critical point the complementary equation can be found by equating Eq. (\ref{eqn:constrained_semi_energy}) to the energy of the paramagnetic ground state $X = 1$, that is 
\begin{equation}
\label{eqn:aux_ground_state}
\sqrt{1 - Z^2} + \frac{W(p)}{p}Z^p = 1,
\end{equation}
by substituting Eq. (\ref{eqn:aux_spinodal_2}) in the above equation and solving for $Z$, then using the found value to solve for $W$, we find the solutions
\begin{equation}
\label{eqn:gsqpt_solutions}
Z_{\rm GS} = \sqrt{\frac{p(p-2)}{(p-1)^2}}, \enspace W_{\rm GS}(p) = \frac{(p-1)^{p-1}}{\sqrt{(p(p-2))^{p-2}}},
\end{equation}
which is the same value in Eq. \ref{eqn:criti_point} in the main text.

The complementary equation in the case of the dynamical critical point is obtained by equating Eq. (\ref{eqn:constrained_semi_energy}) to the energy of the DQPT initial condition $Z = 1$, that is 
\begin{equation}
\label{eqn:aux_dqpt}
\frac{p}{W(p)}\sqrt{1 - Z^2} + Z^p = 1,
\end{equation}
substituting Eq. (\ref{eqn:aux_spinodal_2}) in the above expression, we find the algebraic equation for the $z$-position of the DQPT critical point
\begin{equation}
\label{eqn:algebraic_for_dqpt}
Z^p - \left( \frac{p}{p-1} \right)Z^{p-2} + \frac{1}{p-1} = 0,
\end{equation}
for the models with even value of $p$, $Z = \pm1$ are two solutions of this equation, and for system with odd value of $p$, $Z = 1$ is a solution to this equation. Thus we can make use of these known solutions and reduce the degree of the algebraic equation, allowing us to solve it analytically up to $p=6$. We find for $p=3$
\begin{equation}
Z_{\rm DQPT} = \frac{\sqrt{3}-1}{2}, \enspace W_{\rm DQPT} = \frac{2\sqrt{2}}{(\sqrt{3} - 1)\sqrt{\sqrt{3}}},   
\end{equation}
for $p=4$
\begin{equation}
Z_{\rm DQPT} = \frac{1}{\sqrt{3}}, \enspace W_{\rm DQPT} = \frac{3\sqrt{3}}{\sqrt{2}},   
\end{equation}
for $p=5$
\begin{subequations}
\begin{align}
Z_{\rm DQPT} &= \frac{1}{4}\left( \sqrt{5 + 4\sqrt{5}} - 1\right), \\
W_{\rm DQPT} &= \frac{128\sqrt{2}}{(5 + 4\sqrt{5})^{3/2} \sqrt{5 - 2\sqrt{2} + \sqrt{5 + 4\sqrt{5}}}},
\end{align}
\end{subequations}
and for $p=6$
\begin{subequations}
\begin{align}
Z_{\rm DQPT} &= \frac{1}{\sqrt{10}}\sqrt{1 + \sqrt{21}}, \\
W_{\rm DQPT} &= \frac{50\sqrt{10}}{(11 + \sqrt{21})\sqrt{9 - \sqrt{21}}}.
\end{align}
\end{subequations}
For values of $p>6$ one can then solve Eq. (\ref{eqn:algebraic_for_dqpt}) numerically and find the ``exact'' value of the DQPT critical point. We offer here a way of estimating this critical point which becomes exact in the limit $p\gg1$. We are going to study the properties of Eq. (\ref{eqn:algebraic_for_dqpt}), to do so we define 
\begin{equation}
\label{eqn:}
h(Z;p) = Z^p - \left( \frac{p}{p-1} \right)Z^{p-2} + \frac{1}{p-1},
\end{equation}
notice that $h(0,p)>0$ for all $p\ge2$, and $h(1,p) = 0$ thus $Z = 1$ is always a root. Now, if there is a value $Z_{\rm min}$, $0<Z_{\rm min}<1$, such that $h(Z_{\rm min}, p)$ is a minimum and $h(Z_{\rm min}, p) < 0$, then there is always another value $Z_{\rm DQPT}$, $0<Z_{\rm DQPT}<1$, which is a root of $h(Z;p)$, and monotonicity of $h(Z;p)$ in the interval $Z\in[0,Z_{\rm min}]$ implies that $0<Z_{\rm DQPT}<Z_{\rm min}<1$. 

Now, the extreme points of $h(Z;p)$ are solutions of 
\begin{equation}
Z^{p-3}\left( pZ^2 - p \frac{p-2}{p-1} \right) = 0,    
\end{equation}
thus the point $Z = 0$ is a extreme value only if $p>3$, the other extreme values have the form $Z^* = \pm \sqrt{\frac{p-2}{p-1}}$. The latter gives $\frac{d^2h}{dZ^2}\left.\right|_{Z^*} > 0$, \textit{i.e} it is always a minimum.

Given that $Z = 0$ is a maximum or a saddle, and the point $Z^*$ is always a minimum, it is true that there is a point $Z_{\rm infle}$, the inflection point, such that $0<Z_{\rm infle}<Z^*<1$, at which the concavity of $h(Z;p)$ changes. This point is a solution of $\frac{d^2h}{dZ^2} = 0$ and is given by 
\begin{equation}
\label{eqn:infle_point}
Z_{\rm infle} = \sqrt{\frac{(p-2)(p-3)}{(p-1)^2}},
\end{equation}
and $h(Z_{\rm infle};p)>0$, thus we have $0<Z_{\rm infle}<Z_{\rm DQPT}<Z^*<1$, that is the inflection point is a lower bound to $Z_{\rm DQPT}$.

Finally, we know that $h(Z;p) < 0$ for all $Z\in[Z_{\rm DQPT}, 1]$, with a minimum at $Z^*$. If $h(Z;p)$ were to be symmetric around $Z^*$ in this interval, then the rood $Z_{\rm DQPT}$ will be given by $Z_{\rm DQPT} = 1 - 2(1 - Z^*)$. However, an inspection of $h(Z;p)$ reveals its asymetric character in the interval of interest. Thus, the above value is not the desired root but some approximation to it, we denote it $Z_{\rm DQPT}^*$
\begin{equation}
Z_{\rm DQPT}^* = \frac{2\sqrt{p-2} - \sqrt{p-1}}{\sqrt{p-1}},    
\end{equation}
and $h(Z_{\rm DQPT}^*;p) > 0$, thus this constitutes an upper bound to the desired root $Z_{\rm DQPT}$. Collecting the different points, We have the following chain of inequalities 
\begin{equation}
0 < Z_{\rm infle} < Z_{\rm DQPT} < Z_{\rm DQPT}^* < Z^* < 1,    
\end{equation}
from here we approximate the $z$-position of the DQPT critical point as the arithmetic mean between the lower and upper bounds, $Z_{\rm DQPT} \approx \frac{1}{2}(Z_{\rm infle} + Z_{\rm DQPT}^*)$, giving 
\begin{equation}
\label{eqn:z_dqpt}
Z_{\rm DQPT} \approx \frac{2\sqrt{(p-1)(p-2)} + \sqrt{(p-2)(p-3)} - (p-1)}{2(p-1)},    
\end{equation}
which is true if $p>6$. We notice that the error between the above expression and the numerical solution of Eq. (\ref{eqn:algebraic_for_dqpt}) is $\sim 10^{-3}$ for $p=7$ and decreases from there as a function of $p$. From here we obtain $W_{\rm DQPT}(p)$ as 
\begin{equation}
W_{\rm DQPT}(p>6) \approx \frac{1}{Z_{\rm DQPT}^{p-2}\sqrt{1 - Z_{\rm DQPT}^2}},    
\end{equation}
where $Z_{\rm DQPT}$ is given in Eq. (\ref{eqn:z_dqpt}).

\section{Resonant Hamiltonian and emergent symmetries}
\label{app:resonant_hami_symmetries}
In this Appendix we present the details of the derivation of the resonant Hamiltonian in Eq. (\ref{eqn:resonance_hamiltonian}). We also present the details of the derivation of resonant Hamiltonian associated with the area preserving map in the vicinity of the poles. 

For the resonant Hamiltonian in Eq. (\ref{eqn:resonance_hamiltonian}), the starting point is the delta-kicked Hamiltonian 
\begin{equation}
\label{eqn:app_kick_hamil}
\hat{H}(t) = -\alpha_{\rm B} \hat{S}_x - \frac{\Lambda}{pS^{p-1}}\hat{S}_z^p \sum_{n = -\infty}^\infty \delta (t - n).    
\end{equation}
We will work at a resonance condition, that is, we take $\alpha_{\rm B} = \frac{2\pi}{q}$ with $q$ an integer. As mentioned in Sec. \ref{subsec:resonance_conditions}, at a resonance condition the delta-kicked Hamiltonian can be brought into the form $\hat{H}(t) = \hat{H}_{\rm reso} + \hat{V}(t)$. In order to show this we go to the frame rotating with the precession part of Eq. (\ref{eqn:app_kick_hamil}), that is, interaction picture with respect to $-\alpha_{\rm B}\hat{S}_x$. We then have 
\begin{equation}
\label{eqn:interaction_picture_hamil}
\hat{\tilde{H}}(t) = -\frac{\Lambda}{pS^{p-1}}\left( \cos(\alpha_{\rm B} t)\hat{S}_z - \sin(\alpha_{\rm B} t)\hat{S}_y\right)^p \sum_{n = -\infty}^{\infty}\delta(t - n). 
\end{equation}

At a resonance condition, the system has undergone a full precession after exactly $q$ time steps. So, let us count kicks in groups of $q$, i.e.
\begin{equation}
\label{eqn:dirac_comb_step1}
\sum_{n = -\infty}^{\infty}\delta(t - n) = \sum_{j = 1}^q \sum_{l = -\infty}^\infty \delta\left( t - (lq + j)\right).
\end{equation}
The idea is then to replace the kicking at every step with $ \tilde{H}(t)$ with a kicking every $q$ steps with some other Hamiltonian $\tilde{H}_q(t)$. The main contribution for this operator will be given by
\begin{equation}
     \tilde{H}_q(m) =  \int\limits_{(m-1)q}^{mq} \tilde{H}(t)dt.
\end{equation}
We now compute the integral using that $\alpha_B=2\pi/q$
\begin{multline}
     \int\limits_{(m-1)q}^{mq} (\cos(\alpha_B t)\hat{S}_z-\sin(\alpha_B t)\hat{S}_y)^p \delta(t-lq-j) dt= \\
     \left(\cos\left(\alpha_B(mq+j)\right)\hat{S}_z - \sin\left(\alpha_B(mq+j)\right)\hat{S}_y\right)^p = \\
     \left(\cos\left(\frac{2\pi}{q}j\right)\hat{S}_z - \sin\left(\frac{2\pi}{q}j\right)\hat{S}_y\right)^p
\end{multline}

We then get
\begin{equation}
    \tilde{H}_q(m) \simeq \frac{\Lambda}{p S^{p-1}} \sum\limits_{j=1}^q \left(\hat{\mathbf{O}}_{YZ} \cdot \vec{e}_j \right)^p,
\end{equation}

\noindent which is independent of $m$ up to this order, and where $\hat{\mathbf{O}}_{YZ} = (\hat{S}_y, \hat{S}_z)$ is the projection of the collective spin onto the $y$-$z$ plane, and $\vec{e}_j = \left(-\sin\left(\frac{2\pi}{q}j \right), \cos\left(\frac{2\pi}{q}j \right) \right)$ are the vertices of a $q$-regular polygon. Replacing this into Eq. (\ref{eqn:interaction_picture_hamil}) we have
\begin{equation}
    \tilde{H}(t)\simeq \sum\limits_{m=-\infty}^{\infty} \tilde{H}_q(m) \delta(t-mq).
    \label{eqn:interaction_picture_hamil2}
\end{equation}

Then, we write the Dirac comb defined by the sum over the $m$-index in its Fourier series representation, that is 
\begin{equation}
\label{eqn:dirac_comb_step2}
\sum_{m = -\infty}^\infty \delta(t - mq) = \frac{1}{q}\sum_{m = -\infty}^\infty e^{i\frac{2\pi}{q}m t},
\end{equation}
\noindent and by isolating the term with $m=0$ in the right hand side of Eq. (\ref{eqn:dirac_comb_step2}) we can write
\begin{equation}
\label{eqn:dirac_comb_step3}
\sum_{m = -\infty}^\infty e^{i\frac{2\pi}{q}m t} = 1 + 2\sum_{m = 1}^{\infty}\cos\left(\frac{2\pi}{q}m t\right).    
\end{equation}
Finally, by substituting Eqs. (\ref{eqn:dirac_comb_step1},\ref{eqn:dirac_comb_step2},\ref{eqn:dirac_comb_step3}) into Eq. (\ref{eqn:interaction_picture_hamil2}), we obtain the desired Hamiltonian
\begin{multline}
\label{eqn:full_reso_hamil}
\hat{\tilde{H}}(t) = -\frac{\Lambda}{qpS^{p-1}} \sum_{j = 1}^q \left(\hat{\mathbf{O}}_{YZ} \cdot \vec{e}_j \right)^p \\
- \frac{2\Lambda}{qpS^{p-1}} \sum_{j = 1}^q \left(\hat{\mathbf{O}}_{YZ} \cdot \vec{e}_j \right)^p \sum_{m = 1}^\infty \cos\left(\frac{2\pi}{q}m t \right)
\end{multline}
Hamiltonian in Eq. (\ref{eqn:full_reso_hamil}) is in the desired form, with the resonant Hamiltonian as given in Eq. (\ref{eqn:resonance_hamiltonian}) in the main text.\\

Around the poles, the map in Eq. (\ref{eqn:classical_map}) can be simplified in the following form, consider $Z \to v$ and $Y \to u$ with $u,v \ll 1$, and $X = \pm 1 + \sqrt{u^2 + v^2}\approx \pm 1$, then we have 
\begin{subequations}
\label{eqn:map_around_poles}
\begin{align}
u' &= \left[ u \pm \Lambda v^{p-1} \right]\cos(\alpha(h)) - \sin(\alpha(h))v, \\    
v' &= \left[ u \pm \Lambda v^{p-1} \right]\sin(\alpha(h)) + \cos(\alpha(h))v,
\end{align}
\end{subequations}
we can associated a delta-kicked Hamiltonian to this mapping. It is given by 
\begin{equation}
\label{eqn:delta_hami_poles_1}
\hat{H}(t) = \frac{1}{2}\alpha(h)(u^2 + v^2) \pm \Lambda v^{p-1}\sum_{n = -\infty}^\infty \delta(t - n).
\end{equation}
From now on we will consider the system to be at a resonance condition, that is, we take $\alpha(h) = \alpha_{\rm B} = \frac{2\pi}{q}$ with $q$ and integer. Let us now change variables to polar coordinates, $u = \rho\cos(\varphi)$, $v = -\rho\sin(\varphi)$, with $\rho^2 = u^2 + v^2 \sim 0$ the ``radius of the orbit around the pole", we then re-write Eq. (\ref{eqn:delta_hami_poles_1}) as 
\begin{equation}
\label{eqn:delta_hamil_poles_2}
\hat{H}(t) = \alpha_{\rm B} I \pm (-1)^p \Lambda \left(\rho\sin(\varphi)\right)^p \sum_{n = -\infty}^\infty \delta(t - n),  
\end{equation}
where $I = \frac{1}{2}\rho^2$. We now move to a frame rotating with frequency $\alpha_{\rm B}$, using a canonical transformation generated by the generating function $F = (\varphi - \alpha_{\rm B}t)J$, where the new variables are $\nu = \varphi - \alpha_{B}t$, $J = I$. In this frame the Hamiltonian takes the form 
\begin{equation}
\label{eqn:delta_hamil_poles_3}
\hat{H}(t) = \pm (-1)^p \Lambda \left(\rho \sin(\nu + \alpha_{\rm B}t)\right)^p \sum_{n = -\infty}^\infty \delta(t - n),
\end{equation}
and using the same identities for the dirac-comb in Eqs. (\ref{eqn:dirac_comb_step1},\ref{eqn:dirac_comb_step2},\ref{eqn:dirac_comb_step3}) into Eq. (\ref{eqn:delta_hamil_poles_3}), we obtain 
\begin{multline}
\label{eqn:full_reso_hamil_poles}
\hat{H}(t) = \pm \frac{(-1)^p\Lambda}{q} \sum_{j = 1}^q \left(-\vec{\rho}\cdot\vec{e}_j \right)^p \\
\pm \frac{(-1)^p 2 \Lambda}{q}\sum_{j = 1}^q \left(-\vec{\rho}\cdot\vec{e}_j \right)^p \sum_{m = 1}^{\infty}\cos\left(\frac{2\pi}{q}m(t - j)\right),   
\end{multline}
where $\vec{\rho} = (v, u)$ and $\vec{e}_j = \left(\cos\left(\frac{2\pi }{q}j \right), -\sin\left(\frac{2\pi}{q}j\right) \right)$. Before moving on, let us mention in what sense the terms $\hat{V}(t)$, both in Eqs. (\ref{eqn:full_reso_hamil}, \ref{eqn:full_reso_hamil_poles}), can be treated as a ``small'' perturbation. Although, $\hat{H}_{\rm reso}$ and $\hat{V}(t)$ have the same order of magnitude ($\sim \Lambda$), and thus there is no a small perturbation parameter, the minimal frequency of harmonics in $\hat{V}(t)$ is $q$, if the dynamics driven by $\hat{H}_{\rm reso}$ occurs at a characteristic frequency $\Omega_{q}^{\rm (reso)} = \frac{\Lambda}{q} \ll q$, then $\hat{V}(t)$ can be regarded as a high frequency perturbation and all terms beyond $m\approx 1,2$ can be in principle ignored, were the effect of the terms we kept should be understood as in the principle of averages (see Chapter 3 of~\cite{Wimberger}).

\section{Density of states for the kicked $p$-spin in the vicinity of a point in $\mathcal{B}_{\rm bifu}^{(p)}$}
\label{app:kicked_dos}
In this appendix we use the techniques introduced in Ref~\cite{Bastidas2004} in order to compute the density of states for the kicked $p$-spin model in the vicinity of a point in $\mathcal{B}_{\rm bifu}^{(p)}$. This is done by considering the effective Hamiltonian of $\hat{U}_F^q$.

We focus on the situation of a value of $\alpha_{\rm B} \in \mathcal{B}_{\rm bifu}^{(p)}$ such that the resulting FTC phase has period $q$. Then the Floquet operator of interest is the one driving the dynamics forwards by $q$-steps, that is $\hat{U}_F^q$. As mentioned in Sec.~\ref{subsec:eigenstate_ordering} the effective Hamiltonian, provided $\Lambda$ is small and $\hat{V}(t)$ can be regarded as a high-frequency perturbation, is given by 
\begin{equation}
    \label{eqn:eff_hamil}
    \hat{H}_{\rm eff} = h\hat{S}_x + \frac{\Lambda}{qpS^{p-1}}\sum_{j=1}^q\left(\mathbf{\hat{O}}_{YZ}\cdot \vec{e}_j \right)^p,
\end{equation}
with where $\hat{\mathbf{O}}_{YZ}$, $\vec{e}_j$ defined as in Eq. (\ref{eqn:full_reso_hamil}). The semiclassical energy associated with the effective Hamiltonian in the limit $s\to\infty$ and defining classical variables as $\mathbf{X} = \langle \gamma|\frac{\mathbf{\hat{S}}_{\rm eff}}{S}|\gamma\rangle$, with $|\gamma\rangle$ a spin coherent state, is given by 
\begin{equation}
\label{eqn:classical_energy_q}
    E(\gamma,\gamma^*) = \langle \gamma|\frac{\hat{H}_{\rm eff}}{S}|\gamma\rangle = hX + \frac{\Lambda}{qp}\sum_{j=1}^q\left(\mathbf{O}_{YZ}\cdot \vec{e}_j \right)^p,
\end{equation}
with $\mathbf{O}_{YZ} = (Y,Z)$, and $\gamma$ the complex number obtained via the stereographic projection of the unit vector $(X,Y,Z)$ on the surface of the sphere.

The density of states associated with this Floquet operator is given by $\rho^{(q)}(\epsilon) = \frac{1}{N+1} \sum_\mu \delta \left( \epsilon - \epsilon^{(q)} \right)$. We can write this density of states as 
\begin{equation}
\label{eqn:dos_1}
\rho^{(q)}(\epsilon) = \frac{1}{2\pi(N+1)}\sum_{n = -\infty}^\infty \left( \sum_\mu e^{-in\epsilon^{(q)}} \right)e^{in\epsilon},
\end{equation}
where the term inside brackets are traces of powers of $\hat{U}_F^q$, that is 
\begin{eqnarray}
\left( \sum_\mu e^{-in\epsilon^{(q)}} \right) &=& {\rm Tr} \left[ \sum_\mu e^{-in\epsilon^{(q)}} |\mu\rangle\langle\mu|\right]\nonumber \\
&=& {\rm Tr}\left[(\hat{U}_F^q)^n\right] = {\rm Tr}[\hat{Q}^n],    
\end{eqnarray}
where we have introduced $\hat{Q} = \hat{U}_F^q$. Leading to the following expression for the density of states 
\begin{equation}
\label{eqn:dos_2}
\rho^{(q)}(\epsilon) = \frac{1}{2\pi(N+1)}\sum_{n = -\infty}^\infty {\rm Tr}\left[ \hat{Q}^n \right]e^{in\epsilon},
\end{equation}
the traces can be approximated with the help of the effective Hamiltonian in Eq. (\ref{eqn:eff_hamil}), one has~\cite{Bastidas2004}
\begin{equation}
\label{eqn:traces_1}
{\rm Tr}\left[ \hat{Q}^n \right] = \frac{N+1}{\pi}\int \frac{d^2\gamma}{(1 + \gamma\gamma^*)^2} e^{-i n S E(\gamma, \gamma^*)},
\end{equation}
where $E(\gamma, \gamma^*)$ is given in Eq. (\ref{eqn:classical_energy_q}). The integral can be approximated via stationary phase formula~\cite{Bastidas2004}, one finds 
\begin{equation}
\label{eqn:traces_2}
{\rm Tr}\left[ \hat{Q}^n \right] = \frac{\pi}{N+1}\sum_{\tau \in \mathcal{T}_{\rm st}} \frac{2\pi(1 + \tau\tau^*)^2 e^{i\beta_{\tau}\frac{\pi}{4}} e^{-i n S E(\tau, \tau^*)}}{n S \sqrt{\left|{\rm det}\left[ \mathrm{H}E(\gamma,\gamma^*) \right]\right|_{\gamma = \tau}}},
\end{equation}
where $\mathcal{T}_{\rm st}$ is the set of stationary points of the phase space flow associated with the effective Hamiltonian in Eq. (\ref{eqn:eff_hamil}), $\mathrm{H}E(\gamma,\gamma^*)$ is the Hessian matrix of the classical energy, and $\beta_\tau$ is the index of the stationary point $\tau$, \textit{i.e} the difference between the number of positive and negative eigenvalues of the Hessian.

Using the result in Eq. (\ref{eqn:traces_2}) we can write the density of states as 
\begin{multline}
\label{eqn:dos_3}
\rho^{(q)}(\epsilon) = \frac{1}{2\pi} \\ 
+ \frac{1}{\pi}{\rm Re}\left[ \sum_{\tau\in\mathcal{T}_{\rm st}} \frac{(1 + \tau\tau^*)^2 e^{i\beta_{\tau}\frac{\pi}{4}}} { S \sqrt{\left|{\rm det}\left[ \mathrm{H}E(\gamma,\gamma^*) \right]\right|_{\gamma = \tau}}} \mathrm{Li}_1\left[ e^{i(\epsilon - S E(\tau, \tau^*))} \right] \right],
\end{multline}
with the polylogarithm givne by $\mathrm{Li}_1\left[ e^{i(\epsilon - S E(\tau, \tau^*))} \right] = \sum_{n=1}^\infty\frac{e^{i(\epsilon - s E(\tau,\tau^*))}}{n}$.

With this form of the density of states one can see that at a saddle point, with index equal to zero, there is a logarithmic divergence, see for instance Ref.~\cite{Bastidas2004}. This is the landmark of a clustering ESQPT in systems with one degree of freedom~\cite{Cejnar2021}, leading to the conclusion presented in the main text. Eigenstate clustering in a $q$T-FTC phase, can be diagnosed as an ESQPT of the effective Hamiltonian associated with the $q$-th power of the Floquet operator $\hat{U}_F^q$. 

\section{Absence of higher period FTCs in systems without enough multi-body interactions}
\label{app:absence_FTC}
As discussed in Sec. \ref{sec:reso_reso_cond_bifus} one could use every resonance condition as a point where an FTC phase emerges. However, in such situations the system fails to exhibit a global $\mathbb{Z}_q$ symmetry, requirement which is key for the definition of an FTC phase~\cite{Else2020}. We can build a better understanding of this phenomena by considering an example. Take the system with $p=2$ at the resonance condition $\alpha_{\rm B} = \frac{2\pi}{4}$, value which is not in $\mathcal{B}^{(2)}_{\rm bifu}$, but is a valid resonance condition.

Therefore the classical system sees the emergence of a $1$:$4$ resonance on the $y$-$z$ plane, whose central periodic orbit has its points on the vertices of a square. Thus, if one takes the initial state $\lvert\psi_0\rangle = |S,S\rangle$, and measures $f_{Z}(t)$, a clear period four subharmonic response will be seen. However, the system lacks a global $\mathbb{Z}_4$ symmetry. In fact, the global symmetry of the system at that resonance condition is a $\mathbb{Z}_2 \times \mathbb{Z}_2$ symmetry.  

This system, as all even kicked $p$-spin, has a $\mathbb{Z}_2$ symmetry inherited from the parity symmetry of the clean model. In the mean field limit, this is manifested as invariance of $\mathcal{A}[\mathbf{X}_l;\alpha(h), \Lambda, p]$ to rotations around the $x$-axis by an angle of $\pi$~\footnote{Notice that this implies invariance of the whole phase space portrait to this type of rotations.}, that is, the classical limit of the operator $e^{i\pi\hat{S}_x}$. Consequently, the map $\mathcal{A}^2[\mathbf{X}_l;\alpha(h), \Lambda, p]$ has this symmetry as well. It was originally showed by Haake~\cite{Haake1987} that at the special value of $\alpha(h) = \frac{2\pi}{4}$ the map $\mathcal{A}^2[\mathbf{X}_l;\frac{2\pi}{4}, \Lambda, p]$ is invariant to rotations around the $y$-axis by an angle of $\pi$, that is, the classical limit of the operator $e^{i\frac{2\pi}{4}\hat{S}_y}$. This additional symmetry is a consequence of the system being a double reversible map, and the fact than in those systems parity symmetry can be constructed as the composition of the two involutions defining time reversal operations~\cite{Haake1987,MunozArias2021}. Therefore, at this special value of $\alpha(h)$ the phase space develops a second $\mathbb{Z}_2$ symmetry, this time along the $y$-axis. The overall symmetry group of phase space is $\mathbb{Z}_2 \times \mathbb{Z}_2$. 

On the contrary, consider the system with $p=4$ at this same resonance condition. We know the system has an emergent global $\mathbb{Z}_4$ symmetry, however this symmetry exists independently of the $\mathbb{Z}_2$ inherited from the $p$-spin Hamiltonian. That is, if we block diagonalize $\hat{U}_F$ with respect to the $\mathbb{Z}_2$ symmetry each of the resulting blocks will exhibit this emergent symmetry, which can be diagnosed with the different metrics introduced in the main text. on the contrary, when we block diagonalize $\hat{U}_F$ corresponding to the $p=2$ system, the period-$4$ response emerging around the value of $\alpha_{\rm B} = \frac{2\pi}{4}$ disappears. It is in this sense that the stabilization of higher period FTC phases requires multi-body interactions with $p>2$.

\bibliography{references}
\end{document}